\documentclass[11pt,3p,authoryear]{elsarticle}
\journal{International Journal of Forecasting}
\biboptions{longnamesfirst}
\usepackage{cite}
\usepackage{graphicx,color}
\usepackage{orcidlink}
\usepackage{textcomp}
\usepackage{algorithm}
\usepackage[algo2e]{algorithm2e} 
\usepackage{pgfgantt}
\usepackage{graphicx}
\usepackage{subcaption}
\usepackage{booktabs}
\usepackage{hyperref}
\usepackage{tabularx}
\usepackage{bbm}
\usepackage{stfloats}
\usepackage{amsmath,amssymb,amsfonts}
\newcommand{\argmin}{\operatornamewithlimits{argmin}}

\newcommand{\red}[1]{#1}

\begin{document}

\begin{frontmatter}

\title{Leveraging Continuous Trade Data for Intraday Imbalance Price Forecasting}

\author[ss]{Klimis Stylpnopoulos\corref{cor}}
\ead{2836589s@student.gla.ac.uk}
\author[ss]{Jethro Browell}
\author[ss]{Janine Illian}
\address[ss]{University of Glasgow}

\author[bb]{Ciaran Gilbert}
\author[bb]{Wayne Jones}
\address[bb]{Shell Research LTD}

\cortext[cor]{Corresponding author}

\begin{abstract}
Forecasts of imbalance prices are essential for participants in short-term electricity markets. This study proposes an approach to integrate trade data from continuous intraday markets into forecasting methods and considers the production of both parametric and non-parametric density forecasts. Extensive feature engineering is required to convert irregular trade data into a suitable tabular structure. Our findings demonstrate that intraday trade data can enhance state-of-the-art forecasts based on quantile regression and LASSO. We also propose a model based on Natural Gradient Boosting to predict the means and variances of a bimodal Gaussian mixture, which offers both improved predictive performance and computational efficiency. In a case study based on multiple years of data from Great Britain, our proposed approach reduced the Pinball Loss of state-of-the-art models by 20\% for forecasts made 30 seconds before gate closure, and by 7.5\% for forecasts made one hour before delivery.

\end{abstract}

\begin{highlights}
    \item A novel method for electricity imbalance price forecasting
    \item We engineer features from the GB intraday continuous market
    \item Intraday features improve 15.5 minute-ahead forecasts by 20\%, and hour-ahead by 7.5\%
    \item Methods for parametric and non-parametric probabilistic forecasting are compared
    \item Parametric methods exhibit competitive performance with less computational effort
\end{highlights}


\end{frontmatter}

\section{INTRODUCTION}
\par The electricity market in Great Britain has undergone substantial transformation due to the increasing integration of renewable energy sources and the ongoing need to ensure security of supply. The reliance on weather-dependent renewables introduces uncertainty into short-term electricity markets. As a result, market participants require accurate forecasts to guide bidding strategies and manage operational and financial risks. The Balancing Mechanism offers a price signal that encourages participants to balance their positions, while the Electricity System Operator intervenes to correct real-time imbalances.

\par Although interest in imbalance price forecasting is growing, the academic literature remains limited. This is noteworthy given the rising costs and uncertainties associated with real-time system balancing. Forecasting imbalance prices is challenging due to factors such as high volatility, heteroscedasticity, and the influence of system length—conditions under which the system is either in surplus or deficit—each corresponding to different pricing regimes.

\par In parallel, the frequency and volume of intraday trading have increased markedly in recent years, driven by market volatility, incentives to self-balance, and the expansion of algorithmic trading. This study investigates the predictive value of trading activity in the final hours before delivery as a means to improve imbalance price forecasting. We are only aware of one published study, \citep{Narajewski2022ProbabilisticPrices} that incorporates features derived from the German intraday market.

\par We develop probabilistic forecasting models that quantify predictive uncertainty and compare the performance of state-of-the-art methods, both with and without predictors, derived from intraday trading data. Two modelling frameworks are employed: one uses quantile regression to construct non-parametric predictive densities; the other models the imbalance price as a mixture of parametric densities, weighted by the estimated probability of the system being in surplus or deficit.

\par To evaluate these approaches, we assess the accuracy and calibration of the resulting forecasts, incorporating a comprehensive set of covariates published in real time by the system operator, alongside day-ahead and intraday market data from Great Britain. The main contributions of this work are: demonstration that intraday trade features enhance imbalance price forecast performance of both parametric and non-parametric density forecasts; and, a novel parametric density forecasting method based on a mixture of two Gaussian densities with mixture weights predicted via logistic regression and the mean and variance of the Gaussian densities by  NGBoost \citep{Duan2020NGBoost:Prediction}. The two densities correspond to the long and short regimes of the GB imbalance market, and mixture weights to the predicted probability of each regime being realised. This parametric approach significantly reduces computational cost relative to leading non-parametric methods.


\par The structure of the paper is as follows: Section \ref{sec: literature review} reviews relevant literature on electricity price forecasting. Section \ref{sec: DATA SUMMARY AND ANALYSIS} describes the data sources and key variables. Section \ref{sec: METHODOLOGY} outlines the modelling strategy. Sections \ref{sec: PERFORMANCE METRICS} and \ref{sec: RESULTS} present the evaluation metrics and empirical results, respectively, followed by concluding remarks in Section \ref{sec: CONCLUSION}.

\section{Literature Review}
\label{sec: literature review}

Predicting electricity prices is essential for market participants, and forecast methods are evolving in response to the changing energy system. Uncertainty pervades price forecasting due to imperfect demand forecasts, unexpected outages, and the integration of weather-dependent energy sources. Imbalances between participants' contracted and actual supply/consumption are settled at a price designed to incentivize behaviours that reduce the system net imbalance. Consequently, traders must have good forecasts of imbalance prices.
While day-ahead forecasting has received considerable attention in the academic literature, as reviewed in \citet{Nowotarski2018RecentForecasting, Hong2020EnergyOutlook}, imbalance price forecasting has received relatively little attention.

\par Early works, including \citet{Pinson2007TradingPower}, showed accurate wind forecasts can reduce regulation costs and increase revenue, provided a suitable estimate of imbalance prices is available.
In \citet{Goodarzi2019ThePrices}, the authors explore the relationship between forecast errors from solar and wind energy and the imbalance volume, as well as their effect on the electricity spot price. They utilized Ordinary Least Squares and Linear Quantile Regression to explore the tails of imbalance volume and spot price distribution, and found that renewable energy forecast errors have discernible impact on spot prices. Additionally, \citet{Narajewski2021OptimalCosts} presented trading decisions assuming that participants have optimized their portfolios, utilising econometric models. While \citet{Browell2018RiskMarket} proposed a trading strategy that hedges against penalizing balancing prices in favour of rewarding ones using time series models. 

\par Furthermore, Long-Short-Term-Memory neural networks and Boosting Trees for imbalance forecasting are presented in \citet{Bottieau2020Very-Short-TermSettlement,Lucas2020PriceRegression}. In the former, the authors show that such methods surpass standard benchmark methods; however, they consider the imbalance volume rather than the imbalance price. In the latter, the authors do not compare their methods with standard benchmarks or produce forecasts with uncertainty information (probabilistic forecasts). More sophisticated machine learning techniques are also considered in \citet{Andrade2017ProbabilisticModel}, where high accuracy forecasts are presented for spot and intraday markets. Very-short-term probabilistic forecasting is included in \citet{Narajewski2022ProbabilisticPrices}, where the authors utilise least absolute shrinkage and selection operator (LASSO) to predict the imbalance price, incorporating information from the intraday market along with covariates such as load, renewable energy forecasts, and temporal effects. However, information such as unavailable capacity or intraday-updated forecasts are not included in \citet{Andrade2017ProbabilisticModel, Narajewski2022ProbabilisticPrices}.

\par An additional perspective on imbalance price forecasting is offered in \citet{Browell2022PredictingOpportunities}. Here, forecasts are made at both day-ahead and intraday time scales, where more data from the real-time market are incorporated when it become available as the period being forecasted approaches. Their proposed method consists of a forecasting model that combines forecasts based on the system length\footnote{System length refers to the electricity grid status being in surplus or deficit, designated as long or short, respectively}. In their modelling process, we observe that the bimodal behaviour of imbalance price can be explained with Kernel Density Estimation models in a day-ahead timescale. Although, as the target period approaches, \citep{Browell2022PredictingOpportunities} Linear Quantile Regression is used to forecast the imbalance price, leveraging additional data that becomes available on intraday timescales. Despite mentioning a range of available features that can be used, \citet{Browell2022PredictingOpportunities} do not provide the precise set of co-variates used in their model and other key details are omitted.

\par Expanding on this notion, \citet{Bunn2021AnalysisMarket} incorporate the system length in their forecasting approach, based on the assumption that system length influences the behaviour of the imbalance price, allowing for the identification of distinct regimes in its time series. This is modelled via Markov Regime Switching combined with linear regression  and econometric models such as generalized autoregressive conditional heteroskedasticity (GARCH). \citet{Bunn2021AnalysisMarket} linked the regime switching process to the probability matrix of a Markov model, as a time-varying parameter that can be calculated by the model itself, \citep{Filardo1998BusinessDurations}. The results indicated that MRSM (Markov Regime Switching Model) outperformed their benchmark models, supporting their initial assumption that system length influences the imbalance price. Despite their forecasting being short-term, typically 1 hour ahead, the majority of co-variates used are features published on a day-ahead scale, with constant values throughout the day cycle. This aspect of their modelling approach is crucial when features like renewable forecast errors or de-rated margin can vary during the day. This study did not consider information from the continuous intraday market, which \citet{Browell2022PredictingOpportunities} suggested would add value.

\par Additionally, probabilistic forecasting quantifies uncertainty, which is valuable for energy traders, particularly for managing risk \citep{Lucas2020PriceRegression}. The absence of a probabilistic aspect in forecasting practices hinders traders from improving their trading strategies. Moreover, \citet{Narajewski2022ProbabilisticPrices} discusses how  intraday updated forecasts of power generation can improve price forecast performance. In contrast, \citet{Browell2022PredictingOpportunities} discusses this improvement by incorporating intraday data into their  methodology. However, our literature search found no publications exploring the value of intraday market data or forecast updates for imbalance price forecasting.

\par Hence, we address these gaps in the literature by developing probabilistic forecasting methodologies for  predicting imbalance prices up to one hour-ahead, taking advantage of intraday market data and updated forecast and system information for the first time.

\par It is important to acknowledge the discrepancy in data availability and transparency between academia and industry, which may be partly responsible for some gaps in the literature: electricity market data can be expensive for academics to purchase, and there are obvious disincentives to publish the details of commercial forecasting tools.

\par Specifically, we develop and compare models for parametric and non-parametric density forecasting, reflecting the behaviour of imbalance price in different regimes. Our modelling approaches are enriched with data from the continuous intraday market, which is produced at irregular time intervals and therefore requires specific feature engineering. The effectiveness of incorporating this data is demonstrated by evaluating forecasting performance at key lead times from 15.5-minutes (30 seconds before gate closure) to one hour before delivery. The performance of each model is evaluated using Mean Pinball Loss, Reliability Deviation and Sharpness.

\section{Data Summary And Analysis}
\label{sec: DATA SUMMARY AND ANALYSIS}

\par Imbalance price is influenced by renewable energy production and demand, including related forecasts and forecast errors, generation capacity and outages, as well as other prices and market indexes. We have collected data from Elexon, who operate the balancing mechanism in Great Britain, including making related data publicly available, and EPEX GB, who operate a day-ahead auction and the only continuous intraday power exchange in Great Britain. EPEX market data is available to purchase. Our case study spans the period 2020 to 2023.
\newpage

\begin{figure}
    \centering
    \includegraphics[width=0.9\linewidth]{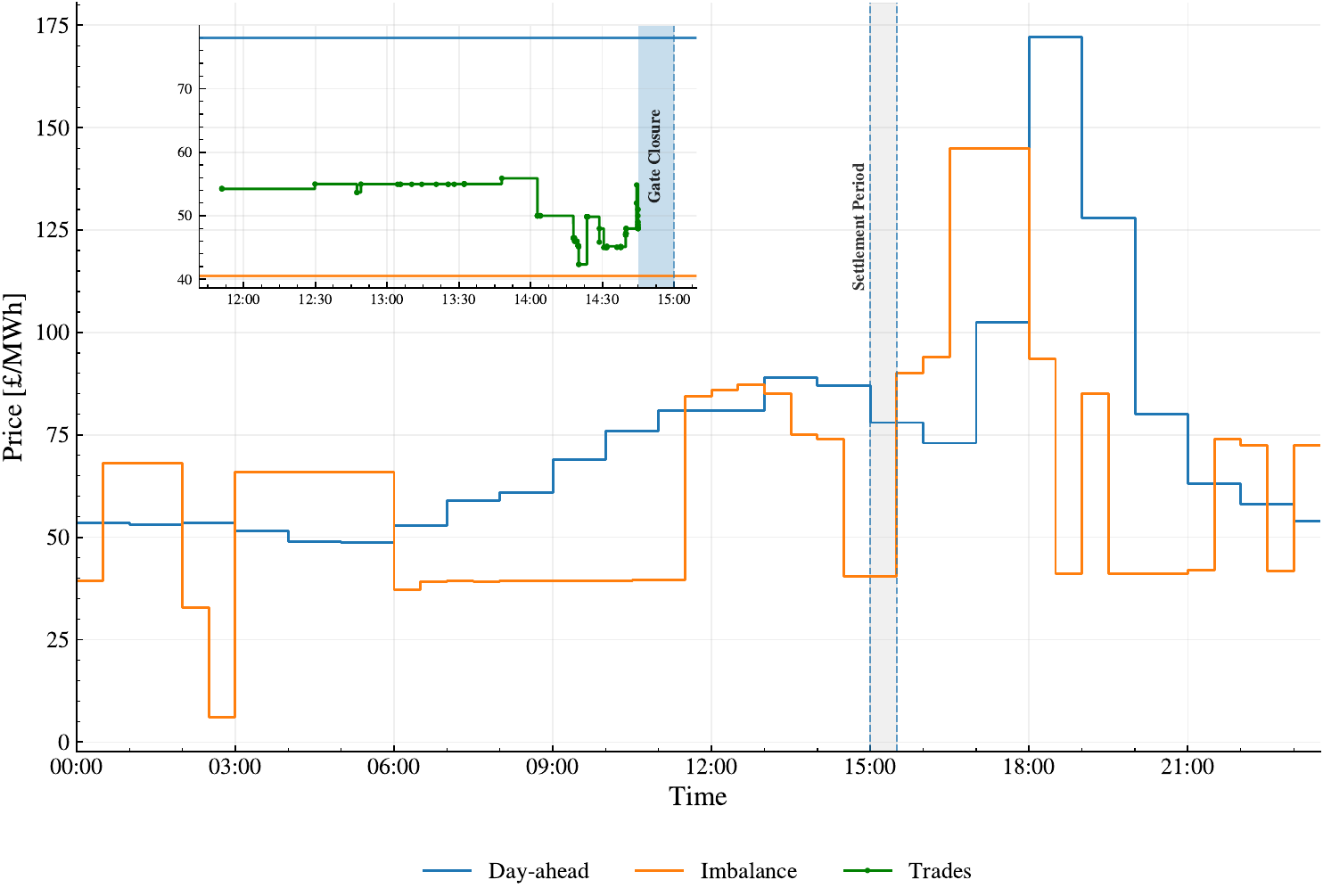}
    \caption{The blue line shows the realized imbalance price, while the orange line reports the corresponding day-ahead price over the same period. \red{The inset displays executed intraday trades (green markers, forward-filled between executions) for a selected delivery period, illustrating high-frequency price formation shortly before gate closure, alongside the corresponding imbalance and day-ahead prices for that settlement period.} The vertical dashed line indicates the settlement period the intraday trades are responding. Period spanning from 2021-01-02 00:00:00 to 2021-01-02 23:30:00, while intraday trades responding to 2021-01-02 15:00:00}
    \label{fig:intra_imba_da_snap}
\end{figure}

\subsection{Balancing Mechanism Data}
\label{sec: BALANCING MECHANISM DATA}
\par Elexon provides API access to real-time and historic data related to balancing and settlement in GB. For this study, we have retrieved generation volumes, generation forecasts, interconnector energy volumes, imbalance volumes, regulation on wholesale energy markets integrity and transparency (REMIT) data, and imbalance prices, as denoted in Table \ref{tab:public_covariates}. Our data collection had a half-hourly frequency and spanned from January 1, 2020, to December 31, 2023.

\begin{table}[H]
\caption{Publicly available features provided by Elexon and National Energy System Operator (NESO). Available features have 30 min frequency. Information about the publicly available features can be found on the \href{https://www.elexon.co.uk/}{Elexon Portal}.}

\centering
\small
\setlength{\tabcolsep}{4pt}
\renewcommand{\arraystretch}{0.9}
\begin{tabularx}{\textwidth}{|l|X|}
\hline
\multicolumn{1}{|c|}{\textbf{Feature}} &
\multicolumn{1}{c|}{\textbf{Details}} \\ \hline

System Sell Price (£/MW) &
Imbalance price calculated without taking into account transmission losses. These values are purely indicative and are used for the settlement of any energy imbalances caused by market participants. \\ \hline

Net Imbalance Volume (NIV) (MW) &
The net imbalance volume is the net sum of all positive and negative system management and energy balancing system actions for each settlement period. \\ \hline

Wind Forecast (MW) &
The wind power production forecast made the night before for each settlement period. \\ \hline

Solar Forecast (MW) &
The solar power production forecast made the night before for each settlement period. \\ \hline

Demand Forecast (MW) &
Forecasted demand provided by the National Grid and published the previous day. \\ \hline

Initial Demand (MW) &
Demand metered by NETSO taking into account transmission losses, but excluding inter-connector transmission losses, pumped storage demand, and station transformer load. \\ \hline

Inter Delta (MW) &
Change in inter-connector import-export netted volumes between subsequent settlement periods summed across six major GB inter-connectors
(IFA\_INT, IFA2\_INT, BRITNED\_INT, NEMO\_INT, MOYLE\_INT, EAST\_WEST\_INT). \\ \hline

Day-Ahead De-rated Margin (MW) &
Difference between the combined generation forecast made 24 hours before the settlement period and the capacity requirement for the market. \\ \hline

Unavailable Capacity Difference (MW) &
Difference between unavailable capacity known one day before and half an hour before, based on REMIT submissions by generators. \\ \hline

NON BM (MW) &
Non-Balancing Mechanism Short Term Operating Reserve volumes instructed by the system operator outside the balancing mechanism. \\ \hline
\end{tabularx}
\label{tab:public_covariates}
\end{table}

\subsection{Market Data}
\label{sec: MARKET DATA}

\par Data from the day-ahead auction and continuous intraday markets operated by EPEX were also collected. The day-ahead auction returns a single price for hourly products, whereas individual bilateral trades of multiple products are traded on the intraday power exchange. The intraday price dataset includes  executed market orders for each product, from which we engineer features to use in imbalance price forecasting models.

\par The data comprising executed orders includes a unique order ID, product, price, volume, execution time, and delivery time. Products for half-hourly, hourly, and 4-hour products are available; however, we focused on the half-hourly product, which is the most liquid. In our data sample, the half-hour power product represents approximately 85\% of all executed trades. Table~\ref{tab:trade_count_summary} reports the number of executed trades retained after applying the selected lead times for the forecasting setup.

\begin{table}[htbp]
\centering
\caption{Summary statistics of trade counts in the time remaining after forecasts are issued. Gate closure (GC) is 15 minutes before delivery.}
\label{tab:trade_count_summary}
\resizebox{\textwidth}{!}{%
\begin{tabular}{lrrrrrrrrr}
\toprule
 Forecast lead-time & Mean & Std. Dev. & Min & P25 & Median & P75 & Max & Zero (\%) \\
\midrule
15.5 min (30 sec to GC)  & 7.45 & 9.62 & 0 & 1 & 4 & 10 & 113 & 18.52 \\
30 min (15 min to GC)   & 95.01 & 74.41 & 0 & 38 & 74 & 135 & 604 & 0.16 \\
60 min (45 min to GC)   & 197.75 & 143.21 & 1 & 84 & 161 & 281 & 1{,}264 & 0.00 \\
\bottomrule
\end{tabular}}
\end{table}

\begin{table}[htbp]
\centering
\caption{\red{Summary statistics of traded volume (MWh) in the time remaining after forecasts are issued. Gate closure (GC) is 15 minutes before delivery.}}
\label{tab:trade_volume_summary}
\resizebox{\textwidth}{!}{%
\begin{tabular}{lrrrrrrrr}
\toprule
 Forecast lead-time & Mean & Std. Dev. & Min & P25 & Median & P75 & Max \\
\midrule
15.5 min (30 sec to GC)  & 20.71 & 25.46 & 0 & 1 & 12 & 30 & 291 \\
 30 min (15 min to GC)   & 216.98 & 113.65 & 0 & 135 & 202 & 283 & 1{,}112 \\
 60 min (45 min to GC)   & 454.17 & 201.92 & 0 & 309 & 434 & 574 & 1{,}852 \\
\bottomrule
\end{tabular}}
\end{table}

\par \red{Tables \ref{tab:trade_count_summary} and \ref{tab:trade_volume_summary} indicate that liquidity in the remaining trading window declines sharply as gate closure approaches. The traded-volume statistics in Table~\ref{tab:trade_volume_summary} make this decline concrete: the median volume available after forecast issuance falls from roughly 434~MWh at the 60-minute lead time to 202~MWh at 30 minutes and to only 12~MWh in the final 30-second window before gate closure--a reduction of more than an order of magnitude. This confirms that even when trades do execute close to gate closure, the volumes involved are small. Notably, for forecasts issued 30 seconds before gate closure, no trades are executed in the remaining window in 18.5\% of settlement periods, meaning that in these cases the forecast could not have been acted upon directly in the continuous market for the target product. Three points are worth emphasising in this regard. First, the absence of \emph{executed} trades does not imply an absence of trading opportunity: resting orders may be present in the order book without being matched, so a participant could still have transacted at posted prices. Second, forecasts issued at this horizon retain decision value beyond trading the target contract itself, informing, for example, the decision to deliberately remain out of balance and settle at the imbalance price, the positioning of physical assets, or bidding in the Balancing Mechanism. Third, all of our models use trades executed in the two hours preceding forecast issuance, so the information content of the intraday features is unaffected by low activity in the final 30 seconds. Nevertheless, the declining liquidity at very short horizons should be borne in mind when interpreting the practical value of the 15.5-minute forecasts.}

\par Furthermore, market orders may be executed at any time, which leads to a dataset without a consistent time frequency. To address this problem, we created price features for each half-hour period $t+k$ that are available at time $t$. For example, the volume weighted average price of trades made between times $t-r_1$ and $t-r_2$ for the half-hour contract with delivery starting at time $t+k$, where $r_1 > r_2 \ge 0$.

\par A commonly used feature in electricity markets is the Volume Weighted Average Price (VWAP) \citep{Narajewski2020EconometricPrices, Janke2019ForecastingTrading}. VWAP combines price and volume over a specified time period. We create 5-minute VWAP in blocks starting from 2 hours before each settlement period, when the most trading activity is observed. As gate closure for each settlement period the continuous intraday market is 15 minutes before delivery starts, the last VWAP block is 4.5 minutes long (20 min to 15.5 min before delivery) so that the corresponding imbalance price forecast can inform decisions in the final 30 seconds of trading. The VWAP in block $b$ for delivery time $t$ is calculated as

\begin{equation}
\label{eq: volume weighted average price}
    Y^{b}_{t} = \frac{\sum_{i=1}^{n}p^{b}_{t,i}u^{b}_{t,i}}{\sum_{i=1}^{n}u^{b}_{t,i}}
\end{equation}
where $i=1,...,n$ indexes individual trades responding to $t$ and $b$, $p_{i,t}^{b}$ is the price of the trade and $u_{i,t}^{b}$ is the volume of trade.

\par Working with the entire distribution of trades can reveal further information not captured by VWAP. Such information informs trading strategy, and therefore impact the imbalance price. In order to represent this trading activity  analytically, we engineer features from the empirical cumulative distribution of the volume-weighted price of trades in 15-minute blocks from 2 hours until just before the gate closure, following \citet{Shinde2021AnalysingMarkets}.  Equation \eqref{eq: Volume-Weighted Empirical Cumulative Distribution Function} presents the Volume-Weighted Empirical Cumulative Distribution Function (VWECDF) normalised by the total volume traded in the given time period
\begin{equation}
\label{eq: Volume-Weighted Empirical Cumulative Distribution Function}
    F_t^{t_{1} \rightarrow t_{2}}(p)=\frac{1}{V^{t_{1} \rightarrow t_{2}}}\sum_{i}\upsilon_{i}\mathbbm{1}_{t_{2}<t\leq t_{1}}
\end{equation}
where $F$ denotes the VWECDF for the half-hour product between the time $t_{1}$ and $t_{2}$ before the settlement period $t$, with $t_{2} > t_{1}$.

\par From $F$ we can create sets $\{ (q_{i}, r_{i}) \}$
where $i$ is the index of trades observed in the time period $t_{1} \rightarrow t_{2}$ ordered by ascending price, $r_{i}$ is empirical frequency (volume) and $q_{i}$ is the empirical quantile (price) after trade $i$. Therefore, we can use linear interpolation to estimate specific quantiles
\begin{equation}
\label{eq: VWECDF linear interpolation}
        q_{\tau}= 
            \begin{cases}
                q_{i},& \text{if } \tau = r_{i}\\
                q_{i} + \frac{q_{i+1}-q_{i}}{r_{i+1}-r_{i}}(\tau - r_{i}),& \text{if } r_{i} <\tau<r_{i+1}
            \end{cases} \quad .
\end{equation}

\par Here, we extract 11 equally spaced quantiles $\tau \in \{0, 0.1,...,1\}$ for each 15-minute block in the two hours preceding the target settlement period
\begin{equation}
\label{eq: VWECDF quantile vector}
    \boldsymbol{q}_{t}^{t_{1} \rightarrow t_{2}} = [ q_{t,0}^{t_{1} \rightarrow t_{2}}, q_{t,0.1}^{t_{1} \rightarrow t_{2}}, ..., q_{t,1}^{t_{1} \rightarrow t_{2}}] \quad .
\end{equation}

\par Another aspect that provides valuable insights into trading activity is the price trend of executed orders as gate closure approaches. We propose an additional feature that describes the VWAP trend for each settlement period, utilising the VWAP values calculated previously and simple linear regression. We fit a linear regression model of VWAP against time to delivery for each settlement period. We extracted the intercept from models which may be interpreted as a naive forecast of the pseudo VWAP at the start of delivery for each settlement period.

\par We emphasize that the VWAP time blocks are calculated from 2 hours until 30 seconds before the settlement period. Having that said, the estimated price for every settlement period can be calculated from the intercept of every linear regression model we estimated, for each settlement period. 

\par Lastly, we denote the price spread as $P_{t}$, which is our response variable. The price spread is calculated by subtracting the day-ahead price from the imbalance price to remove long-term effects and because both imbalance and intraday prices are best interpreted relative to the price of electricity established in the day-ahead market. For this reason we also subtract the day-ahead price from all engineered price features, including those derived from intraday trades. 

\section{Methodology}
\label{sec: METHODOLOGY}
In this section, we adopt a framework to predict price spreads between the day-ahead and imbalance price. \red{We begin with a simple persistence-based Naive benchmark, before introducing} feature sets used by all models, followed by linear quantile regression for non-parametric density forecasting, followed by Mixture Models for parametric density forecasting with two regimes. LASSO is used in both cases, and an additional tree-based method, Natural Gradient Boosting (NGBoost), is used to estimate a more flexible version of the Mixture Model. \red{The procedure used to select the LASSO shrinkage parameter $\lambda$, and the criterion optimised in each case, is described in Section~\ref{sec: RESULTS}.}

\begingroup
\subsection{Naive Benchmark}
\label{sec: naive benchmark}
As a simple lower-bound benchmark, we consider a naive persistence forecast for the price spread $P_{t}$ of the target settlement period. The point forecast is given by the most recently available realised price spread at the time the forecast is issued,
\begin{equation}
\label{eq: naive point forecast}
    \hat{P}_{t} = P_{t^{*}},
\end{equation}
where $t^{*}$ denotes the settlement period whose realised price spread is the most recent available at the forecast issuance time. Because realised imbalance prices become available only after a reporting lag following the end of each settlement period, and forecasts are issued at different lead times $k$ before delivery, the settlement period $t^{*}$--and hence the age of the persisted observation--differs across lead times: the longer the lead time, the older the most recently available observation. \red{We account for this reporting lag when selecting $t^{*}$, so that only settlement periods whose price spread would genuinely be available at the forecast issuance time are used. We do not, however, model the fact that published imbalance prices are themselves subject to revision as subsequent settlement runs are calculated: the value of $P_{t^{*}}$ is taken from the single price-spread series used consistently throughout this study, rather than from the specific (potentially since-revised) vintage that would have been observable in real time. This simplification, common in the electricity price forecasting literature, is applied only to the choice of which value to persist and does not otherwise affect the construction of the target variable or of any other model.} To obtain a probabilistic forecast, we construct quantile predictions by adding the empirical quantiles of the in-sample persistence errors to the point forecast,
\begin{equation}
\label{eq: naive quantile forecast}
    \hat{P}^{\tau}_{t} = \hat{P}_{t} + \hat{e}^{\tau}, \qquad \hat{e}^{\tau} = \text{the } \tau\text{-th empirical quantile of } \{P_{i} - \hat{P}_{i}\}_{i=1}^{T},
\end{equation}
where $\{P_{i}-\hat{P}_{i}\}_{i=1}^{T}$ are the persistence errors observed on the training set. This benchmark uses no explanatory variables and serves as a lower bound against which the added value of the remaining models, all of which incorporate publicly available and/or intraday market information, can be assessed.
\endgroup

\par We consider that price spread can be explained by intraday trading price features, energy forecasts and weekday dummies. We define the vector $\boldsymbol{x}_{t}$ of explanatory variables and $\boldsymbol{\beta}$ a vector that contains their corresponding coefficients and construct a set of linear predictors as \red{(here and throughout, $\boldsymbol{x}_{t}$ and $\boldsymbol{\beta}$ are both column vectors of matching dimension, and we write $\boldsymbol{x}_{t}^{\top}\boldsymbol{\beta}$ for their inner product, which is a scalar; this convention is used consistently in \eqref{eq: prediction model} and \eqref{eq: qr regression vectors})}

\begin{align}
\label{eq: prediction model}
    \boldsymbol{x}_{t}^{\top}\boldsymbol{\beta} = &\underbrace{\sum_{j \in S}\sum_{i \in C}\delta_{j,i}Y_{t+k+j|t}}_\textrm{VWP from previous, future and current contracts} + \nonumber \\ & \underbrace{\sum_{l=1}^{4}\sum_{b=0}^{1}\nu_{l,b}Y^{b}_{t+k|t}}_\textrm{VWP spread in quantiles} + 
            \underbrace{\omega Y_{t+k|t}^{intercept}}_\textrm{VWP intercept} + \nonumber \\ 
            & \underbrace{\sum_{h=1}^{9}\gamma_{h}Z_{h, t+k|t}}_\textrm{Publicly available predictors} + \underbrace{\sum_{m=0}^{48}\zeta_{m}D_{m,t}}_\textrm{Settlement period dummies} + \nonumber \\& \underbrace{\sum_{g \in A}\xi_{g}R_{g,t}}_\textrm{Informative Missing Predictors}
\end{align}
where $t$ is the delivery time for the target settlement period, $k$ is the forecast lead-time. We incorporate trades from all contracts that are available at the time we conduct the forecast, in a 2-hour window. Let $Y_{t+k+j|t}$ denote the volume weighted price spread information available at time $t$, encompassing data from previous, future and current contracts. Let $S$ be a dynamically populated set representing the number of contracts (previous, future, and current) in our model, and $C$ a set containing the volume-weighted price spread in 5-minute blocks within a 2-hour window, both contingent upon the forecasting horizon. With $Y_{t+k|t}^{b}$ we denote the volume weighted price spread in quantiles, with $b$ denoting the quantile level of the empirical distribution of volume weighted price. Also, with $Z$ we denote the publicly available features and with $D$ a dummy variable which indicates the settlement period. Lastly, we may have missing values in our intraday features, due to absence of trades in any time window. We handle this issue by creating responding vectors for every intraday feature that will inform our model if there are executed trades or not, \citep{Groenwold2020InformativeKnowing}: Let $A$ be a dynamically populated set containing the number of price features our model utilises, and let $R$ be the categorical variable indicating whether there is an order at settlement period $t$ (0 for no orders and 1 for order presence).

\par \red{Formally, let $n_{g,t}$ denote the number of executed trades in the time window associated with intraday feature $g \in A$ for target settlement period $t$. The missingness indicator is defined as}
\begin{equation}
\label{eq: informative missingness}
\red{
    R_{g,t} = \mathbbm{1}\{n_{g,t} > 0\} =
    \begin{cases}
        1, & \text{if at least one trade is executed in the window of feature } g\\
        0, & \text{otherwise}
    \end{cases}}
\end{equation}
\red{\noindent so that each intraday price feature is accompanied by its own indicator. When $R_{g,t}=0$, the corresponding price feature is undefined and is imputed with zero, which corresponds to the day-ahead price after the price-spread transformation described in Section~\ref{sec: MARKET DATA}. The key idea of \emph{informative missingness} \citep{Groenwold2020InformativeKnowing} is that the absence of trades is not missing at random: periods without trading activity typically coincide with low liquidity or an absence of new information reaching the market, both of which are themselves informative about the imbalance price. Including $R_{g,t}$ as an additional regressor allows the model to learn a systematic effect associated with the absence of trades, via the coefficients $\xi_{g}$ in \eqref{eq: prediction model}, while preventing the imputed values from biasing the coefficients of the observed price features.}

\subsection{Linear Quantile Regression}
\label{sec: Linear Quantile Regression}
\par Quantile regression, first introduced by \citet{Koenker1978RegressionQUANTILES}, differs from ordinary least squares regression by focusing on estimating coefficients for various quantiles rather than just the mean of the dataset. One of its key advantages is that it does not rely on assumptions about the data sample’s distribution. This method allows for the examination of relationships between dependent and independent variables across different quantiles, offering a detailed understanding of how independent variables impact the dependent variable across various price ranges.

\par Importantly, quantile regression can provide valuable information for our dependent variable in the low and extreme quantiles of our price range. Therefore, this modelling approach can help energy traders evaluate price risks and develop bidding strategies.

\par Let $\tau \in (0,1)$ indicate the probability level, hence we present our quantile regression model in \eqref{eq: qr regression vectors}

\begin{equation}
\label{eq: qr regression vectors}
    Q_{\tau}(P_{t}|\boldsymbol{x}_{t}) = \boldsymbol{x}_{t}^{\top}\boldsymbol{\beta^{\tau}}
\end{equation}

\par Given the large number of features from the intraday market, we use $l_{1}$-penalised quantile regression, \citep{Uniejewski2021RegularizedForecasting, Koenker1978RegressionQUANTILES}. This approach shrinks the coefficients of features that do not enhance the model's loss function, such as the quantile loss, towards zero. \red{We adopt the idea of combining $\ell_{1}$-penalisation with quantile regression from \citet{Uniejewski2021RegularizedForecasting}, who introduce LASSO-penalised Quantile Regression Averaging (LQRA). We note an important methodological distinction from their approach: LQRA applies quantile regression, with LASSO-based shrinkage, to a set of \emph{point forecasts produced by several individual models}, thereby combining (averaging) those forecasts into a single probabilistic forecast. In contrast, our QR-LASSO applies $l_{1}$-penalised quantile regression directly to the vector of explanatory features $\boldsymbol{x}_{t}$ described in Section~\ref{sec: METHODOLOGY}, and does not combine forecasts from multiple models. We therefore retain the name QR-LASSO, rather than LQRA, to avoid suggesting that our method performs forecast averaging.} As presented in \eqref{eq: qr regression vectors}, we find the quantile regression coefficients $\boldsymbol{\beta}^\tau$ from the minimization problem in

\begin{align}
\label{eq: optimisation for quantile regression}
    \hat{\beta}^{\tau} = \argmin_{\beta^{\tau}} &\Biggl [ ~\sum_{t=1}^{T}\underbrace{(\tau-\mathbbm{1}_{P_{t} <Q_{\tau}(P_{t}|\boldsymbol{x}_{t})})(P_{t}-Q_{\tau}(P_{t}|\boldsymbol{x}_{t}))}_\textrm{quantile loss function} + \underbrace{\lambda\sum_{n=1}^{N}|\beta_{n}^{\tau}|}_\textrm{$l_{1}$-norm penalisation} \Biggr ]
\end{align}

\noindent In our chosen quantile space, we trained 19 models—one for each quantile—and then generated our quantile predictions for the price spread, denoted as $\hat{P}^{\tau}_{t}$. As mentioned above, $l_{1}$-norm penalization is applied to the quantile regression model, denoted as \textit{QR-LASSO}, which incorporates data from the intraday market. However, our benchmark model, denoted as \textit{QR-Benchmark}, does not contain any features from the intraday data; therefore, penalization is not present. Consequently, the optimization problem for our benchmark will follow the equation \eqref{eq: optimisation for quantile regression}  with $\lambda$=0.

\par \red{Explicitly, QR-Benchmark is the specialisation of the general linear predictor in \eqref{eq: prediction model} obtained by removing all terms derived from intraday trade data (the volume-weighted price terms, the volume-weighted price quantile terms, the volume-weighted price intercept term, and the informative-missingness indicators, all of which are undefined in the absence of intraday data), leaving only the publicly available predictors and the settlement-period dummies:
\begin{equation}
\label{eq: qr benchmark formula}
    \boldsymbol{x}_{t}^{\top}\boldsymbol{\beta} = \underbrace{\sum_{h=1}^{9}\gamma_{h}Z_{h, t+k|t}}_\textrm{Publicly available predictors} + \underbrace{\sum_{m=0}^{48}\zeta_{m}D_{m,t}}_\textrm{Settlement period dummies}.
\end{equation}
The quantile predictions of QR-Benchmark are then obtained by substituting \eqref{eq: qr benchmark formula} into \eqref{eq: qr regression vectors}, with the optimisation problem in \eqref{eq: optimisation for quantile regression} solved with $\lambda=0$, as no intraday-derived features are present to penalise.}

\subsection{Parametric Approach}
\label{sec: Parametric Approach}

\par
\begin{figure}
\centerline{\includegraphics{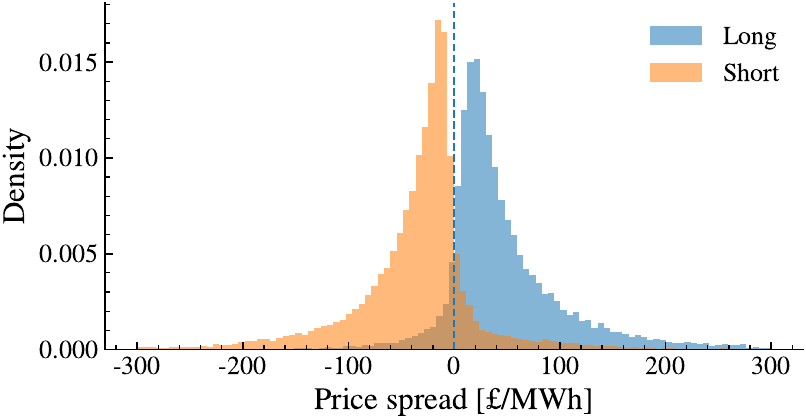}}
\caption{Histogram of the marginal price spread over the period from 1 January 2020 to 31 December 2023, with shaded regions indicating whether the system is in deficit or surplus. Extreme values are excluded.}
\label{fig: price spread distribution}
\end{figure}

 In this section, our study utilizes data from the GB electricity market, spanning multiple years from the GB Balancing Mechanism, to develop a parsimonious model; this predicts the imbalance price (relative to the known day-ahead price) from just before until two settlement periods ahead of delivery, on a rolling half-hourly basis. We introduce methods for forecasting the price spread as mixture parametric distributions. To account for the bimodal nature of our target variable, we employ a parametric model characterized by a mixture of two Gaussian distributions, Figure \ref{fig: price spread distribution}

\begin{equation}
\label{eq: gaussian densities}
    p(y|\boldsymbol{x}) = \sum_{i=1}^{s}w_{i}(\boldsymbol{x})\psi(y|\mu_{i}(\boldsymbol{x}), \sigma_{i}(\boldsymbol{x}))
\end{equation}
where $s=2$ is the number of components in the mixture, $\boldsymbol{x}$ is the vector of features, $w_{i}(X)$ is the weight assigned to each density, $\mu_{i}(X)$ is the mean value of each density and $\sigma_{i}(X)$ is the standard deviation of each density. 

\par In the following sections, we present modelling approaches that incorporate linear regression and tree-based algorithms to estimate the parameters of our predicted distribution. We utilise the entire training set, but conditionally filter the data based on whether the system is in surplus or deficit. This allows us to estimate the model parameters accordingly. The calibrated models are then applied to the same testing dataset from which we derive the predicted mean, $\mu_{i}(\boldsymbol{x})$, and standard deviation, $\sigma_{i}(\boldsymbol{x})$.

\par The weights $w_s(\boldsymbol{x}_t)$ represent the conditional probabilities of the system being in state $s$
and are estimated using a logistic regression model, which will be introduced in the next section. The final
equation is presented below

\begin{equation}
\label{eq: conditional predicted distribution}
    \hat{f}(P_{t}|\boldsymbol{x}_{t}) = \sum_{s=0}^{1}\hat{w}_{s}(\boldsymbol{x}_{t})
    \psi(P_{t}|\hat{\mu}_{s}(\boldsymbol{x}_{t}), \hat{\sigma}_{s}(\boldsymbol{x}_{t}))
\end{equation}

\noindent where $s \in \{0,1\}$ denotes the system state (long or short), $t$ is the settlement period,
$\hat{w}_{s}(\boldsymbol{x}_{t})$ is the predicted probability of the system being in state $s$,
and $\psi(P_{t}|\hat{\mu}_{s}(\boldsymbol{x}_{t}), \hat{\sigma}_{s}(\boldsymbol{x}_{t}))$ is the
predicted conditional density with mean $\hat{\mu}_{s}(\boldsymbol{x}_{t})$ and standard deviation
$\hat{\sigma}_{s}(\boldsymbol{x}_{t})$.

\subsection{Predicting the System Length}
\label{sec: logistic regression}

\par In this section, we present a binary classification model to predict whether the grid status is in deficit or surplus. The Net Imbalance Volume (NIV) is encoded as a binary indicator, taking the value 1 when the grid is in deficit and 0 when it is in surplus. All remaining features coincide with those used in \eqref{eq: qr regression vectors}. We employ a Logistic Regression model with the linear combination of our explanatory variables, as outlined in \eqref{eq: prediction model}

\par Lastly, we utilise Logistic Regression in order to predict the probability of the system status at the respective forecasting horizon $k$. Below, we provide the general function of Logistic Regression

\begin{equation}
\label{eq: log reg}
    \pi(\boldsymbol{x}_{t}) = \frac{1}{1+e^{-( \boldsymbol{x}_{t}^{\top}\boldsymbol{\beta})}}
\end{equation}
where $\pi(\boldsymbol{x}_{t})$ denotes the probability of the system being in the long state, where $\boldsymbol{x}{t}$ is the vector of explanatory variables and $\boldsymbol{\beta}$ the associated coefficient vector. Parameter estimation is performed by maximising the log-likelihood augmented with an $\ell_{1}$--penalty (LASSO) to induce sparsity in the coefficients.

We predict the probability of the system being in short state with $1 - \pi(\boldsymbol{x}_{t})$. The predicted probabilities of the system status being long or short are used as weighing components $\hat{w}_{s}(\boldsymbol{x}_{t})$ with $s \in \{0,1\}$ in \eqref{eq: conditional predicted distribution} as shown below

\begin{equation}
\label{eq: logreg weighing components}
    \hat{w}_{s}(\boldsymbol{x}_{t}) = \begin{cases}
        \pi(\boldsymbol{x}_{t}) & \text{ for $s=1$} \\
        1-\pi(\boldsymbol{x}_{t}) & \text{ for $s=0$} \quad \text{.}
    \end{cases}
\end{equation}

\subsection{Lasso Mixture Model}
\label{subsec: LASSO MIXTURE MODEL}

\noindent In our first parametric modelling approach, called as \textit{MM-LASSO}, we utilise LASSO to estimate the parameters of our predicted distribution. Initially, we construct our linear model conditional on the system status $s$ being long or short, utilising \eqref{eq: prediction model}, as outlined in \eqref{eq: lasso mixture model}

\begin{equation}
    \label{eq: lasso mixture model}
    \hat{\mu_{s}} = \boldsymbol{x}_{t}^{\top}\boldsymbol{\beta}_{s}
\end{equation}

\noindent where $s \in \{0,1\}$. With \eqref{eq: lasso mixture model} we find the  $\boldsymbol{\beta}_{s}$ from the minimization problem provided below:

\begin{align}
    \label{eq: conditional lasso estimator}
    \widehat{\boldsymbol{\beta} }_{s}=&\argmin_{\boldsymbol{\beta}_{s} \in \mathbbm{R}^{n}}
    \sum ^{T}_{t=1, t \in \{t|s_{t}=s\}}\left( \mu_{s} -\boldsymbol{x}_{t}^{\top}\boldsymbol{\beta}_{s}\right) ^{2} + \lambda _{s}\sum ^{n}_{i=1}\left| \boldsymbol{\beta}_{s,i}\right|
\end{align}

\noindent where $T$ is the size of our training set, $n$ is the size of our coefficient vector, and $\lambda_{s}$ is the lasso estimator. Furthermore, for, $\lambda_{s}=0$ the equation \eqref{eq: conditional lasso estimator} takes the form of Ordinary Least Squares. On the other hand, if $\lambda_{s}$ is extremely large then $\widehat{\beta}_{s}=0$, shrinking all our parameter coefficients to zero.

\par We calibrate our model using our training set conditional to the system status being long or short, in order to estimate the model coefficients. We employ our calibrated models conditional on system status in order to predict the mean using the test set. Additionally, we calculate the residuals from our conditional training sets in order to calculate the conditional standard deviation. Here, the standard deviation has a fixed value in each condition. We present the Gaussian densities conditional to the system being in state $s$ in \eqref{eq: lasso cases}

\begin{equation}
\label{eq: lasso cases}
\psi(P_{t}|\hat{\mu}_{s},
    \hat{\sigma}_{s}) = 
    \begin{cases}
        \psi(P_{t}|\hat{\mu}_{0},\hat{\sigma_{0}}) & \text{for $s = 0$} \\
        \psi(P_{t}|\hat{\mu}_{1},\hat{\sigma_{1}}) & \text{for $s=1$}
    \end{cases}  
\end{equation}

\noindent where $s \in \{0,1\}$ is the system being short and long accordingly, $\hat{\mu_{s}}(\boldsymbol{x}_{t})$ is the predicted mean conditional on the system status and $\hat{\sigma_{s}}$ is the predicted fixed standard deviation, calculated from the in-sample residuals. Below, we present the equation we utilise to calculate the variance from in-sample residuals

\begin{equation}
\label{eq: lasso mixture standard deviation}
        \hat{\sigma}^{2}_{s} = \frac{\sum_{t \in \{t|s_{t}=s\}}(P_{t} - \hat{\mu}_{s})^{2}}{|\{t|s_{t}=s\}|}
\end{equation}

\noindent where $s \in \{0,1\}$ is the system status being short and long accordingly. With \eqref{eq: lasso cases}, \eqref{eq: lasso mixture standard deviation} and \eqref{eq: logreg weighing components} we have our predicted distribution in the form as \eqref{eq: conditional predicted distribution}.

\par In a later approach, we present another method utilising LASSO, named \textit{LR-LASSO}. In this approach, we assume that the predicted distribution follows a normal distribution with a fixed predicted variance calculated from the in-sample residuals. The design of this model follows the presented model in Section \ref{subsec: LASSO MIXTURE MODEL} but without filtering the dataset by the system status $s$. We anticipate that this approach will demonstrate the effect of fixed variance of the predicted distribution. 

\begingroup
\subsection{Historical Simulation LASSO}
\label{subsec: HS-LASSO}

We additionally introduce \textit{HS-LASSO}, a benchmark that uses the same point forecast as LR-LASSO but replaces the parametric Gaussian assumption with a nonparametric, historical-simulation-style treatment of the residuals. This isolates the effect of the distributional assumption from the point-forecasting model itself, since both LR-LASSO and HS-LASSO share an identical conditional-mean forecast $\hat{\mu}_{t} = \boldsymbol{x}_{t}^{\top}\hat{\boldsymbol{\beta}}$, estimated exactly as in Section~\ref{sec: Parametric Approach} but without conditioning on system status.

Rather than assuming Gaussian residuals, quantile forecasts are constructed by adding the empirical quantiles of the in-sample residuals to the point forecast:
\begin{equation}
\label{eq: hs-lasso quantiles}
    \hat{P}^{\tau}_{t} = \hat{\mu}_{t} + \hat{e}^{\tau}, \qquad \hat{e}^{\tau} = \text{the } \tau\text{-th empirical quantile of } \{P_{i} - \hat{\mu}_{i}\}_{i=1}^{T},
\end{equation}
where $\{P_{i}-\hat{\mu}_{i}\}_{i=1}^{T}$ are the in-sample residuals of the LASSO point forecast. This is the same historical-simulation construction used for the Naive benchmark (Section~\ref{sec: naive benchmark}), applied here to a regression-based point forecast rather than a persistence forecast. Unlike LR-LASSO, which forces every quantile onto a symmetric distribution around $\hat{\mu}_{t}$ regardless of the true shape of the residual distribution, HS-LASSO imposes no distributional assumption and can therefore accommodate skewness or other departures from normality in the residuals, at the cost of assuming that the empirical distribution of in-sample residuals is representative of out-of-sample forecast errors.
\endgroup

\subsection{NGBoost Mixture Model}
\label{subsec: NGBOOST MIXTURE MODEL}

\par In a different modelling approach, NGBoost offers the flexibility to predict both the mean and the standard deviation of the predicted distribution. This algorithm assumes that $P_{\theta}(y|x)$ follows a normal distribution, where $x$ represents a vector of independent variables and $y$ is the target variable. The model estimates $p$ parameters of $\theta \in \mathbb{R}^{p}$ for the distribution as functions of $x$. NGBoost leverages the Gradient Boosting method, as detailed in \citet{Ke2017LightGBM:Tree}, utilising Decision Trees as base learners and a proper scoring rule to refine the outputs at each iteration 
\par As a proper scoring rule, \citet{Gneiting2007StrictlyEstimation} use the parameters of the forecast distribution and an outcome to assign a score to the forecast, ensuring that the true distribution achieves the best expected score, as shown in \eqref{eq: proper scoring rule}.

\begin{equation}
    \label{eq: proper scoring rule}
    \mathbb{E}_{y|Q}[S(Q,y)] \leq \mathbb{E}_{y|Q}[S(P,y)] , \quad \forall Q,P
\end{equation}

\par NGBoost employs gradient descent, denoted as $\triangledown S(\theta, y)$, on a scoring rule over a parameterised probability distribution $P_{\theta}$, with $\theta$ as the parameter and $y$ as the outcome, here, we use the CRPS. However, this gradient is not invariant to re-parameterisation, meaning that the distance between two parameter values does not necessarily reflect the distance between the two distributions represented by those parameters. To address this, NGBoost incorporates the Natural Gradient into its boosting algorithm. \citet{Duan2020NGBoost:Prediction} explain that divergences induced by $D_{KL}$, \citep{Kullback1951OnSufficiency} and $D_{L^2}$, \citep{Machete2013ContrastingRules} are invariant to how $P$ and $Q$ are parameterised. Additionally, they note that small changes in parameters correspond to points on a statistical manifold, with each point representing a probability distribution. 

The Natural Gradient then takes the form of \eqref{eq: natural gradient}

\begin{equation}
    \label{eq: natural gradient}
    \Tilde{\nabla}S\left( \theta ,y\right) \propto I_{s}\left( \theta \right) ^{-1}\nabla S\left( \theta ,y\right)
\end{equation}

\noindent where $I_{s}\left( \theta \right)$ is the Riemannian metric of the
statistical manifold at $\theta$, influenced by our selected proper scoring rule $S$. Additionally, they provide individual forms of $I_{s}\left( \theta \right)$, for example when Maximum Likelihood Estimation (MLE) as proper scoring rule, then $I_{s}\left( \theta \right)$ is the Fischer information or when using the CRPS score the same is the Riemannian metric of the statistical manifold that uses $D_{L^{2}}$ as local distance. In summary, $I_{s}\left( \theta \right)$ provides more information about the local geometry of the statistical manifold around the parameter $\theta$. \red{We use the Natural Gradient Boosting algorithm exactly as proposed by \citet{Duan2020NGBoost:Prediction}, without modification; Algorithm~\ref{alg:ngboost} reproduces their procedure for completeness and to fix notation used in the remainder of the paper.} In Algorithm \ref{alg:ngboost} we present the steps of NGBoost

\begin{algorithm}
  \SetAlgoLined
  \SetKwInOut{Initialize}{Initialize}
  \SetKwInOut{Input}{Input}
  \SetKwInOut{Output}{Output}
  \SetKwInOut{Data}{Data}

  \caption{\red{Natural Gradient Boosting Algorithm, reproduced from \citet{Duan2020NGBoost:Prediction}}}
  \label{alg:ngboost}
  \Data{%
  Dataset $D=(\boldsymbol{x}_{t}, P_{t})_{t \in \{t|s_t = s\}}$
  }
  
  \Input{%
    $M$: number of iterations \\
    $\eta$: learning rate \\
    $\theta$: probability distribution parameters \\
    $S$: Proper scoring rule CRPS \\
    $f$: Decision Trees as base learner
    
  }
  
  \Initialize{%
    Model with a constant value for $\theta_{0} = \argmin_{\theta}\sum^{n}_{i=1}S(y_{i}, \theta)$
  }
  
  \For{$m=1$ \KwTo $M$}{
    \For{$i=1$ \KwTo $n$}{
    $r^{m}_{i} =  I_{S}(\theta^{(m-1)}_{i})^{-1}\triangledown_{\theta}S(y_{i},\theta^{(m-1)}_{i})$
    }
    
    Fit $\{x_{i},r_{i}^{m}\}^{n}_{i=1}$ to $f^{(m)}$ \\
    $\rho^{m}=\argmin_{\rho}\sum^{n}_{i=1}S(\theta^{m-1}_{i}-\rho \cdot f^{m}(x_{i}),y_{i})$ \\
    \For{$i=1$ \KwTo $n$}{
      Update $\theta^{m}_{i}=\theta^{m-1}_{i}-\eta (\rho^{m}\cdot f^{m}(x_{i}))$
    }
  }
  
  \Output{%
    $\{{\rho^{m}, f^{m}}\}^{M}_{m=1}$
  }
\end{algorithm}

\par Our data input denoted as $(\boldsymbol{x}_{t}, P_{t})_{t \in \{t|s_{t}=s\}}$ where $\boldsymbol{x}_{t}$ is the vector containing our independent variables and $P_{t}$ is our target variable. Our independent variables vector includes: Volume Weighted Price, Volume Weighted Price Quantiles, Volume Weighted Price Intercept, Initial Demand, Wind forecast error, Solar forecast error, Demand error, Inter-connector Imports/Exports, derated margin, REMIT unavailable capacity, Net Imbalance Volume, NON-BM Volume and day of the week indicator. 

\par We calibrate our model, called as \textit{MM-NGBoost}, to every conditional training set, and then we produce $\hat{\mu}_{s}(\boldsymbol{x}_{t})$ and $\hat{\sigma}_{s}(\boldsymbol{x}_{t})$ using our test set. In contrast to the methodologies discussed in Section \ref{subsec: LASSO MIXTURE MODEL}, where the estimation of $\hat{\sigma}_{s}(\boldsymbol{x}_{t})$ is derived from the in-sample residuals and has a fixed value, with NGBoost, we are able to estimate the standard deviation of our Gaussian densities conditional on the system being in the state $s$ for every settlement period $t$. Therefore, we present our predicted Gaussian densities conditional for the system being in state $s$ in \eqref{eq: ngboost cases}

\begin{align}
\label{eq: ngboost cases}
\psi(P_{t}|\hat{\mu}_{s},
    \hat{\sigma}_{s}) = 
    \begin{cases}
        \psi(P_{t}|\hat{\mu}_{0},\hat{\sigma}_{0}) & \text{for $s = 0$} \\
        \psi(P_{t}|\hat{\mu}_{1},\hat{\sigma}_{1}) & \text{for $s=1$}
    \end{cases}  
\end{align}

Additionally, using \eqref{eq: ngboost cases} and  \eqref{eq: logreg weighing components}, we construct our predictive distribution using \eqref{eq: conditional predicted distribution}.

\begin{table}[htbp]
\centering
\caption{\red{Summary of the point-forecast and probabilistic forecasting components of each model considered in this study. ``Intraday data'' indicates whether the model's feature set includes covariates derived from continuous intraday trades.}}
\label{tab:model summary}
\begin{tabular}{p{2.1cm}p{4cm}p{4.7cm}c}
\toprule
Model & Point / conditional-mean forecast & Probabilistic forecast & Intraday data \\
\midrule
Naive & Persistence (last observable price spread) & Empirical quantiles of in-sample persistence errors & No \\
QR-Benchmark & -- & Linear quantile regression (\eqref{eq: qr regression vectors}--\eqref{eq: optimisation for quantile regression}, $\lambda=0$) on public fundamentals only & No \\
QR-LASSO & -- & $l_{1}$-penalised linear quantile regression (\eqref{eq: qr regression vectors}--\eqref{eq: optimisation for quantile regression}) & Yes \\
LR-LASSO & LASSO-penalised linear regression (conditional mean) & $\mathcal{N}(\hat{\mu}, \hat{\sigma})$, fixed $\hat{\sigma}$ from in-sample residuals & Yes \\
HS-LASSO & Same point forecast as LR-LASSO & Empirical quantiles of in-sample residuals (historical simulation) & Yes \\
MM-LASSO & LASSO-penalised linear regression, conditional on system regime $s$ & Two-component Gaussian mixture; fixed $\hat{\sigma}_{s}$ per regime from in-sample residuals & Yes \\
MM-NGBoost & NGBoost, conditional on system regime $s$ & Two-component Gaussian mixture; NGBoost predicts both $\hat{\mu}_{s}$ and $\hat{\sigma}_{s}$ (heteroskedastic) & Yes \\
MM-NGBoost-F & Same as MM-NGBoost, trained on fundamentals-only feature set & Same as MM-NGBoost & No \\
\bottomrule
\end{tabular}

\vspace{2pt}
{\footnotesize\red{Note: for MM-LASSO, MM-NGBoost and MM-NGBoost-F, the mixture weights (i.e.\ the probability of the system being long or short) are obtained from a common $l_{1}$-penalised logistic regression classifier (Section~\ref{sec: logistic regression}), estimated on the same feature set as the corresponding mixture component.}}
\end{table}

\section{Performance Metrics}
\label{sec: PERFORMANCE METRICS}
\par In this section, we detail the performance metrics chosen for our case study. As we are producing density forecasts, we evaluate them according to the principle of sharpness subject to calibration. We use Mean Pinball Loss (MPL) to summarise performance, Reliability Deviation to verify calibration and Sharpness (average interval width) to compare the spread or confidence of our forecasts \citep{Pinson2006PropertiesEvaluation,Zhao2020AnGeneration}. We also use skill scores to compare each model against the benchmark model in terms of Mean Pinball Loss.

\subsection{Mean Pinball Loss}
\par Quantile Loss or Pinball Loss $\mathcal{L}_\tau$ is the score corresponding to the quantile $\tau$. It quantifies the absolute error between the actual and predicted values by applying different weights depending on whether the error is positive or negative, as given by

\begin{equation}
\label{eq: quantile loss}
    \mathcal{L}_{\tau}(P_{t}, \hat{P}_{t}) = \begin{cases}
                q|P_{t} - \hat{P}_{t}|, &\text{if } P_{t} > \hat{P}_{t}\\
                (1-q)|P_{t} - \hat{P}_{t}|, &\text{if } P_{t} \leq \hat{P}_{t}
            \end{cases} \quad .
\end{equation}

After we derive the quantile loss for every prediction we made, at every quantile level, we evaluate the accuracy of our probabilistic forecasts by averaging the quantile loss across all quantiles

\begin{equation}
\label{eq: Mean Pinball Loss}
    \Bar{\mathcal{L}}_{\tau} = \frac{1}{T}\sum_{t=1}^{T}\mathcal{L}_{\tau}(P_{t}, \hat{P}_{t})
\end{equation}

\noindent where Mean Pinball Loss, denoted as $\Bar{\mathcal{L}_{\tau}}$, is the quantile loss averaged by probability level.

\subsection{Reliability Deviation}
\par Reliability $\mathcal{R}$ of a probabilistic forecast is a measure that evaluates the degree to which predicted probabilities correspond to the observed relative frequencies in our data \citep{Messner2020EvaluationView}. To assess the reliability of our probabilistic forecasts, we plot the predicted probabilities within the desired interval
$(0,1)$ against the observed relative frequencies. To calculate the observed relative frequency for a specific sub-interval, we aggregate the actual outcomes associated with the probabilistic forecasts within that sub-interval and divide this sum by the total number of forecasts. This process is repeated for each sub-interval. In \eqref{eq: reliability}, we present the equation for reliability at a specific sub-interval, which, in our case, corresponds to the quantile level $q$

\begin{equation}
\label{eq: reliability}
    \mathcal{R}_{\tau} = \frac{1}{T}\sum_{t=1}^{T}\mathbbm{1}(P_{t} \leq \hat{P}^{\tau}_{t})
    \quad .
\end{equation}

\par Ideally, our objective is to ensure that the observed relative frequencies closely align with the predicted probabilities. As previously discussed, under these conditions, the reliability plot would exhibit a diagonal line. Therefore, instead of evaluating the reliability of every model against the diagonal, we measure the Reliability Deviation ($\mathcal{R}^{dev}$) of every model, against the diagonal

\begin{equation}
    \mathcal{R}^{dev}_{\tau} = R_{\tau} - \tau \quad .
\end{equation}

\subsection{Sharpness}
\par We evaluate the sharpness of our probabilistic forecasts, calculating the average width of our prediction intervals. A prediction interval with nominal level $\alpha$, incorporating all the available information at time $t$ in a pair of quantiles as defined in \eqref{eq: PI}

\begin{equation}
\label{eq: PI}
    \hat{\mathcal{I}}_{t+k|t}^{\alpha} = [Q_{t+k|t}^{\underline{\tau}}, Q_{t+k|t}^{\overline{\tau}}]
\end{equation}

where $\overline{\tau} - \underline{\tau} = \alpha$.

Therefore, we present the sharpness ($\Bar{\mathcal{I}^{\alpha}}$) or average prediction interval width as defined in \eqref{eq: sharpness}

\begin{equation}
\label{eq: sharpness}
    \Bar{\mathcal{I}}^{\alpha} = \frac{1}{T}\sum_{t=1}^{T}(Q_{t+k|t}^{\overline{\tau}} - Q_{t+k|t}^{\underline{\tau}}) \quad .
\end{equation}

\subsection{Skill Score}
\par In forecasting practices, it is common to compare different models against a chosen benchmark using a metric tailored to specific requirements. In our case, we aim to evaluate the accuracy of each model by employing the Mean Pinball loss at key lead times relative to our benchmark. The equation for calculating the Skill Score of a selected model in comparison to our benchmark is provided in \eqref{eq: skill score}

\begin{equation}
\label{eq: skill score}
    SkillScore = \frac{\Bar{\mathcal{L}}_{benchmark} - \Bar{\mathcal{L}}_{model}}{\Bar{\mathcal{L}}_{benchmark}}
\end{equation}

\par The Skill Score is a metric used to assess each model's accuracy relative to our benchmark at key lead times. In this study, we compare the forecasting performance of several models that incorporate information from the intraday electricity market with that of a selected benchmark that does not. Our objective is to quantify the accuracy improvements of each model across different lead times.

\section{Model Estimation}
\label{sec: RESULTS}
We test the performance of each model in two different scenarios: static and batch. In the static scenario, we use the period from January 1, 2020 to December 31, 2021, to train our models; January 1, 2022, to December 31, 2022 for validation; and January 1, 2023 to December 31, 2023 for out-of-sample testing to evaluate final forecast performance.

\par In the batch scenario, we re-train each model on a monthly basis throughout the test set using hyper-parameters selected in the static set-up. The initial training set is from January 1, 2020 until December 31, 2022 and models are used to forecast throughout January 2023. Models are then re-trained using data from January 1, 2020 to January 31, 2023 and used to forecast through February 2023, and so on. However, we do not re-tune model hyper-parameters at every iteration due to the high computational cost this would entail.

\par For the LR-LASSO and MM-LASSO we use the sliding window method to select the shrinkage parameter $\lambda$. We calculate the mean squared error for each validation fold. Our validation set comprises 12 validation folds, one per month for the entire year of 2022. Therefore select the $\lambda$ parameter that is one standard deviation from the minimum mean squared error, to select a conservative level of regularisation. We also use BayesianOptimizationCV by \citet{Head2022Scikit-optimize/scikit-optimize}
to tune the optimal $\lambda$ parameter in the QR-LASSO model, following the same training/validation procedure described above, \red{except that, since QR-LASSO is estimated by quantile regression rather than ordinary least squares, $\lambda$ is selected to minimise the quantile (pinball) loss \eqref{eq: Mean Pinball Loss} on the validation folds, in place of the mean squared error used for LR-LASSO and MM-LASSO.}

\par Hyper-parameters of the MM-NBoost models are selected in a single train-validate set-up which consists of two stages. The first goal is to define the number of boosting trees for Algorithm \ref{alg:ngboost}. We set the learning rate to 0.05 and calculate the CRPS score between our training and validation sets at each iteration. Additionally, the iteration process stops if there is no gain in the CRPS score for 20 consecutive iterations.

\par In the second stage, fixing the boosting trees and learning rate at the values from stage one, we optimise the maximum tree depth, minimum samples per leaf, and minimum samples per split using Tree-structured Parzen Estimator implemented in Optuna framework \citep{Akiba2019Optuna}. We perform 100 trials with different hyper-parameter combinations, optimising the CRPS score using our validation set. This procedure is followed for each lead-time and system length, six models in total.

\par \red{We adopt this two-stage design for two reasons. First, in gradient boosting the number of boosting iterations and the learning rate together govern the overall model capacity and interact strongly with one another, so that the optimal number of trees is highly sensitive to the learning rate. Fixing the learning rate and determining the number of trees by early stopping on the validation CRPS is the standard and computationally efficient way to resolve this interaction, since it identifies the appropriate model complexity in a single training run rather than treating the number of trees as a free parameter to be searched over. Second, jointly searching over the number of trees together with the tree-structure hyper-parameters would enlarge the search space substantially and make the optimisation both more expensive and more prone to overfitting the validation set. By first fixing the model complexity and then tuning only the tree-structure hyper-parameters that control the bias--variance trade-off of the individual learners, we obtain a smaller and better-conditioned search space in the second stage, which the Tree-structured Parzen Estimator can explore efficiently within a modest number of trials. This staged approach is therefore both computationally cheaper and less susceptible to overfitting than a single joint search, while still tuning all of the hyper-parameters to which the model is most sensitive.}

\section{Forecast Performance}
\label{sec: model performance}
\begin{figure*}
\centering
\centerline{\includegraphics[width=1.1\textwidth]{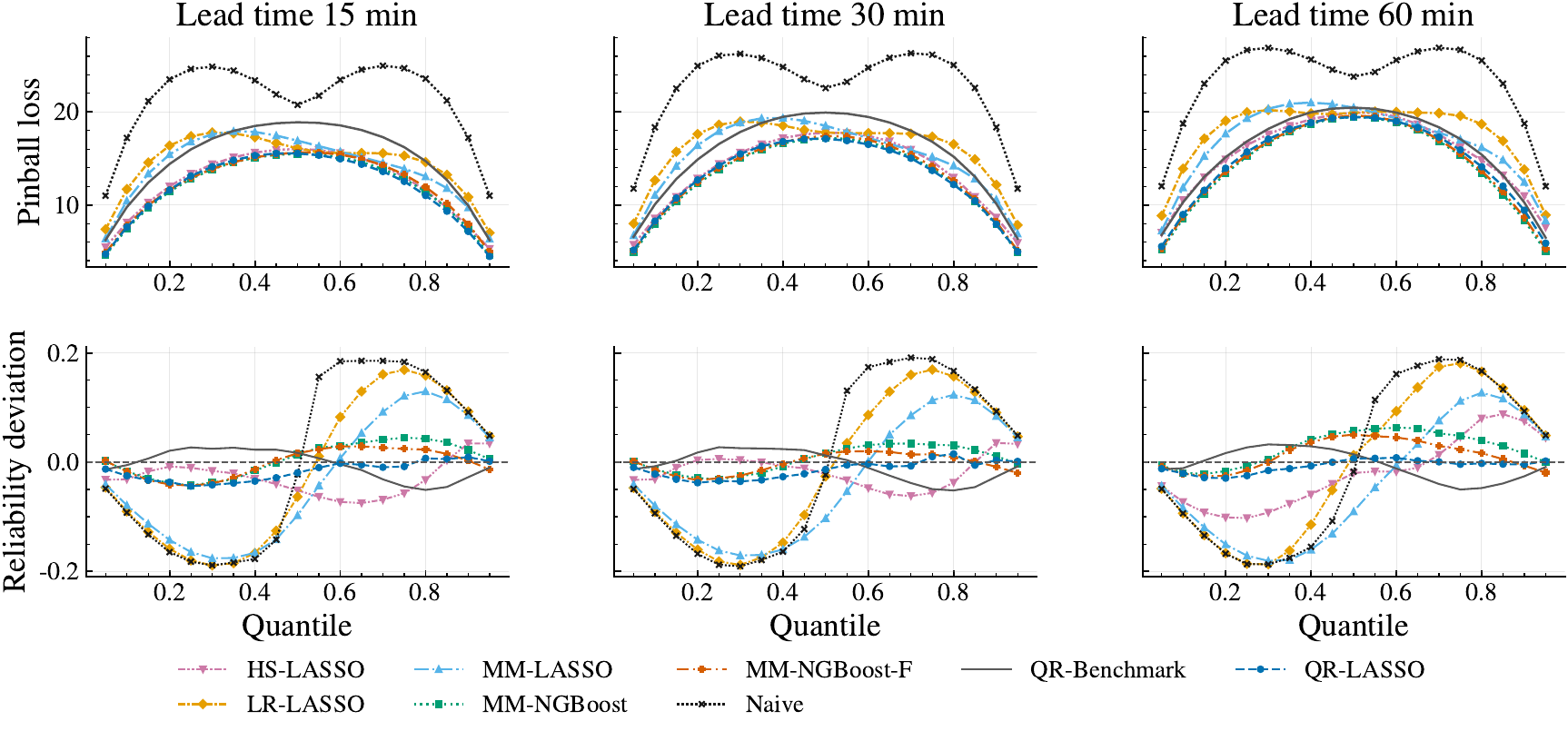}}
\caption{Mean Pinball Loss \eqref{eq: Mean Pinball Loss} and Reliability Deviation \eqref{eq: reliability} at key lead times for each model are evaluated. The results follow a setup where predictions are made in monthly batches from 2023 onward, while the validation set is fixed for the entire year of 2022. We evaluate our probabilistic forecasts using the entire year of 2023. QR-LASSO and MM-NGBoost, which account for heteroskedasticity, show consistent accuracy against the QR-Benchmark for every lead time, while LR-LASSO and MM-LASSO, which both assume constant variance, perform poorly. \red{We also report MM-NGBoost-F, a variant of the proposed mixture model trained on the same fundamentals-only feature set as QR-Benchmark (i.e.\ without intraday trade features), to isolate the contribution of intraday trade data within the mixture-model paradigm; see Section~\ref{sec: model performance} for discussion.}
\label{fig: mpl-rdl results}}
\end{figure*}

In Figure \ref{fig: mpl-rdl results}, we present the Mean Pinball Loss and Reliability Deviation for each model and lead-time. Note all models evaluated for the entire year of 2023.

\par In terms of Pinball, forecasts produced by QR-LASSO and MM-NGBoost outperform QR-Benchmark and other models, and perform similarly to one another for all lead-times. 
The only difference between QR-Benchmark and QR-LASSO is that the latter incorporates information from intraday trading, so we can attribute the significant performance gain of 10\% to 20\% to these new features. \red{This attribution is corroborated by the permutation feature importance analysis reported in Appendix~A: when intraday features are included, several of the volume-weighted price and quantile features derived from the continuous market rank among the most influential predictors, particularly in the tails of the predictive distribution and at longer lead times. The mechanism is intuitive: while QR-Benchmark relies almost entirely on the lagged price spread, the upper and lower edges of the volume-weighted traded-price distribution in the window before gate closure become the dominant predictors for QR-LASSO once intraday features are available, and they do so specifically at the corresponding (highest and lowest) quantiles--that is, the most recent trades are most informative precisely where imbalance-price risk is largest. This explains both why the intraday features yield their largest gains in the tails of the predictive distribution and why those gains are most pronounced at the shortest lead time, when the traded-price information is freshest.}

\par \red{To assess whether this trade-data effect holds across modelling paradigms, we repeat the comparison for the proposed mixture model, introducing MM-NGBoost-F, a version of MM-NGBoost trained on the same fundamentals-only feature set as QR-Benchmark, i.e.\ excluding all covariates derived from intraday trade data. Table~\ref{tab:mm-ngboost-ablation} reports the Mean Pinball Loss Skill Score of each model relative to QR-Benchmark. Two results stand out. First, MM-NGBoost-F, despite using no intraday information whatsoever, already achieves a skill score of 18.5\% at the 15.5-minute lead time, close to the 20.4\% achieved by QR-LASSO with full access to intraday trade data; this shows that a substantial share of the overall improvement over QR-Benchmark is attributable to the mixture model's ability to capture the heteroskedastic and regime-dependent nature of the price spread, rather than to the intraday features themselves. Second, the marginal gain from adding intraday features within the mixture-model paradigm (MM-NGBoost-F $\rightarrow$ MM-NGBoost) is modest, at 1--3 percentage points across lead times, in contrast to the 7--20 percentage point gain observed within the linear quantile-regression paradigm (QR-Benchmark $\rightarrow$ QR-LASSO). Taken together, these results indicate that intraday trade data and model flexibility act as partial substitutes: a sufficiently flexible model can recover much of the same predictability from publicly available fundamentals alone, while intraday data yields its largest incremental value for simpler, linear model classes. Combining both remains the best approach, as MM-NGBoost, which uses both, achieves the highest skill score at every lead time.}

\begin{table}[htbp]
\centering
\caption{\red{Mean Pinball Loss Skill Score (\%) relative to QR-Benchmark, batch training. MM-NGBoost-F is trained on the same fundamentals-only feature set as QR-Benchmark. The marginal gain from intraday features within the mixture-model paradigm (MM-NGBoost-F $\rightarrow$ MM-NGBoost) is small relative to the gain within the linear paradigm (QR-Benchmark $\rightarrow$ QR-LASSO), despite the overall skill of MM-NGBoost being highest throughout.}}
\label{tab:mm-ngboost-ablation}
\begin{tabular}{lrrr}
\toprule
Pinball Skill Score [\%] & 15.5 min & 30 min & 60 min \\
\midrule
QR-Benchmark (no intraday data) & 0.0 & 0.0 & 0.0 \\
QR-LASSO (+ intraday data) & 20.4 & 16.0 & 6.9 \\
MM-NGBoost-F (no intraday data) & 18.5 & 15.2 & 8.3 \\
MM-NGBoost (+ intraday data) & 20.2 & 16.6 & 9.6 \\
\bottomrule
\end{tabular}
\end{table}

\par \red{While the marginal gain from intraday features on Mean Pinball Loss is modest for the mixture model, a complementary picture emerges when the full predictive distribution is examined via the (quantile-based) Continuous Ranked Probability Score. Table~\ref{tab:crps-tail} compares MM-NGBoost and MM-NGBoost-F on the median CRPS, the 95th percentile of CRPS, and the proportion of forecasts exceeding a threshold set at the 90th percentile of the pooled CRPS of the two models--a proxy for the incidence of large forecast errors. MM-NGBoost improves on MM-NGBoost-F consistently across every lead time and on every statistic: the median CRPS is lower, the upper tail of the CRPS distribution is lighter, and the proportion of forecasts falling above the pooled 90th-percentile threshold falls from just over 10\% for MM-NGBoost-F to under 10\% for MM-NGBoost at every lead time. This indicates that, although intraday features contribute only a small reduction to the average pinball loss within the mixture-model paradigm, they consistently reduce the incidence of large forecast errors, which is of particular operational relevance for balancing and trading decisions.}

\begin{table}[htbp]
\centering
\caption{\red{Comparison of MM-NGBoost and MM-NGBoost-F using the (quantile-based) CRPS approximation, batch training. ``\% above threshold'' is the share of forecasts with CRPS exceeding the 90th percentile of the pooled CRPS of the two models at that lead time; under equal performance this would be 10\% for each model.}}
\label{tab:crps-tail}
\begin{tabular}{lrrrrrr}
\toprule
& \multicolumn{2}{c}{15.5 min} & \multicolumn{2}{c}{30 min} & \multicolumn{2}{c}{60 min} \\
\cmidrule(lr){2-3} \cmidrule(lr){4-5} \cmidrule(lr){6-7}
& MM-NGBoost & -F & MM-NGBoost & -F & MM-NGBoost & -F \\
\midrule
Median CRPS       & 16.50 & 16.98 & 17.82 & 18.47 & 20.53 & 20.96 \\
P95 CRPS          & 56.09 & 56.91 & 61.64 & 62.43 & 65.84 & 66.56 \\
\% above threshold & 9.6 & 10.4 & 9.7 & 10.3 & 9.7 & 10.3 \\
\bottomrule
\end{tabular}
\end{table}

\par In contrast, the models MM-LASSO and LR-LASSO exhibited poor performance, even failing to improve in the benchmark for longer lead-times despite access to additional explanatory information. This underperformance is due to the assumption of fixed variance in these models, which hampers their ability to accurately capture the heteroscedasticity of the price spread. The fixed variance assumption leads to less sharp and poorly calibrated forecasts, resulting in higher mean pinball loss values across all lead-times.

\par \red{To assess whether this underperformance stems from the parametric Gaussian assumption itself, rather than from the underlying point forecast, we introduce HS-LASSO (Section~\ref{subsec: HS-LASSO}), which uses the same point forecast as LR-LASSO but constructs quantiles nonparametrically via historical simulation instead of assuming Gaussian residuals. HS-LASSO substantially outperforms LR-LASSO at every lead time: under batch training, HS-LASSO achieves a Mean Pinball Loss Skill Score relative to QR-Benchmark of 17.1\%, 12.3\% and 1.8\% at 15.5, 30 and 60 minutes respectively, compared with 2.2\%, $-3.4\%$ and $-11.9\%$ for LR-LASSO over the same lead times; the pattern is similar under fixed training (17.5\%, 13.8\% and 5.2\% for HS-LASSO versus 0.7\%, $-3.3\%$ and $-21.4\%$ for LR-LASSO). Notably, LR-LASSO's Gaussian assumption is not merely suboptimal but actively harmful at longer lead times, performing \emph{worse} than QR-Benchmark despite using the same point forecast and additional intraday information, whereas HS-LASSO, built on an identical point forecast, remains competitive throughout. This improvement is statistically significant according to the Diebold--Mariano test (Appendix~B) at every lead time and under both the batch and fixed training set-ups. This confirms that the Gaussian assumption, rather than the LASSO point forecast itself, is responsible for LR-LASSO's underperformance: the same point forecast, freed from the assumption of symmetric, fixed-shape residuals, performs considerably better. HS-LASSO also performs significantly better than QR-Benchmark at every lead time, providing a benchmark that, like QR-Benchmark, requires no intraday trade data, but is based on a regression point forecast combined with a standard nonparametric post-processing technique rather than direct quantile regression or persistence.}

\par In terms of reliability, non-parametric models, such as QR-LASSO and QR-Benchmark, and MM-NGBoost, which predicts the variance, are calibrated and exhibit only small deviations from perfect reliability, shown in Figure \ref{fig: mpl-rdl results}. QR-LASSO shows slightly better calibration than MM-NGBoost and QR-Benchmark. The reliability deviation is similar for these three models across all lead-time except for MM-NGBoost, which deteriorates slightly for the 60-minute lead-time.

\par Otherwise, the relative performance of all models is the same across all lead-times, accompanied by an increase in Mean Pinball Loss for greater lead-times. This is something to be expected, as the greater the lead time, the less information is available to inform forecasts.

\par
\begin{figure*}
\centering
\centerline{\includegraphics[width=1.1\textwidth]{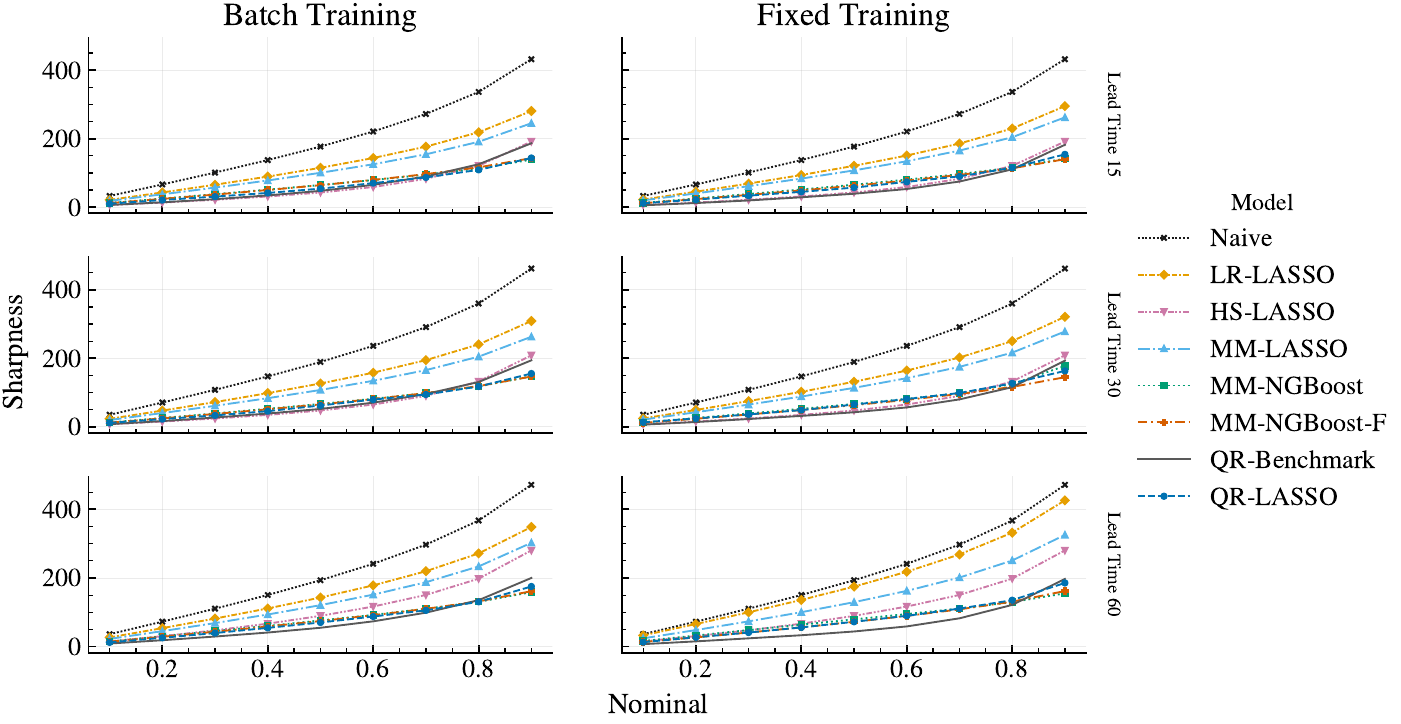}}
\caption{Sharpness \eqref{eq: sharpness} values are evaluated at nominal proportion levels for each model at specific lead times. The first column presents results for models trained in a batch training scenario, while the second column shows results for models trained with a fixed training set. Our top-performing models, QR-LASSO and MM-NGBoost, show narrower prediction intervals in each setup, while LR-LASSO and MM-LASSO produce wider prediction intervals. This highlights the importance of accounting for conditional variance, which results in sharper forecasts. \red{MM-NGBoost-F, the fundamentals-only variant of MM-NGBoost, achieves sharpness close to that of QR-LASSO and MM-NGBoost despite lacking intraday trade features, consistent with the modest marginal contribution of these features to sharpness within the mixture-model paradigm.}\label{fig: sharpness results}}
\end{figure*}

 In Figure \ref{fig: sharpness results} we present the sharpness of the forecasts. As with Mean Pinball Loss, QR-LASSO and MM-NGBoost have similar performance, however the average interval with is similar to that of QR-Benchmark, despite the difference in Pinball. This suggests that while forecasts produced by the benchmark, with no intraday trade data, have similar average interval widths to QR-LASSO and MM-NGBoost, the latter successfully modulate their forecasts to reflect increased or decreased uncertainty, informed by intraday trade data.

\par Consistent with the Mean Pinball Loss, LR-LASSO and MM-LASSO have relatively large average interval widths. This highlights that access to intraday data alone is insufficient to yield forecast improvement, as the gains appear to come from not only improved precision, but also improved discrimination between periods of high and low uncertainty. We also confirm that the batch training setup adds value to forecasting performance for all lead-times. Additionally, the Mean Pinball loss shows a reduction of 5£/Mwh for lead times 15.5 and 30 min and 2£/Mwh for lead time 60 min for QR-LASSO and MM-NGBoost compared to QR-Benchmark.

\par \red{The statistical significance of the performance differences discussed above is assessed formally in Appendix~B using pairwise Diebold--Mariano tests. These confirm that QR-LASSO and MM-NGBoost produce significantly more accurate forecasts than the QR-Benchmark at all lead times, and that the separation between the leading methods and the fixed-variance alternatives becomes more pronounced as the forecast horizon lengthens; at the 60-minute lead time MM-NGBoost emerges as the dominant method. Appendix~C complements the probabilistic evaluation with point-forecast accuracy, reporting MAE and RMSE for the median of each predictive distribution; the ranking of methods is consistent with the Mean Pinball Loss results, with QR-LASSO and MM-NGBoost delivering the most accurate point forecasts overall. One exception is worth noting: LR-LASSO's RMSE becomes competitive with, and at the 60-minute lead time even marginally better than, the leading models, despite its poor Mean Pinball Loss. This is expected rather than contradictory--RMSE is minimised by the conditional mean, and LR-LASSO is the only model in our comparison explicitly optimised for that target, whereas the remaining models target the median (equivalent to minimising MAE) or the full predictive distribution; LR-LASSO's weak probabilistic performance stems specifically from its fixed-variance assumption, not from its point forecast. Finally, Appendix~D repeats the evaluation under a fixed (non-updating) training set. Differences relative to the monthly batch set-up presented here are small, indicating that our conclusions are robust to the choice of training strategy, although batch re-training yields modest additional gains and is closer to operational practice.}

\par
\begin{figure}
\centering
\centerline{\includegraphics[width=1.1\textwidth]{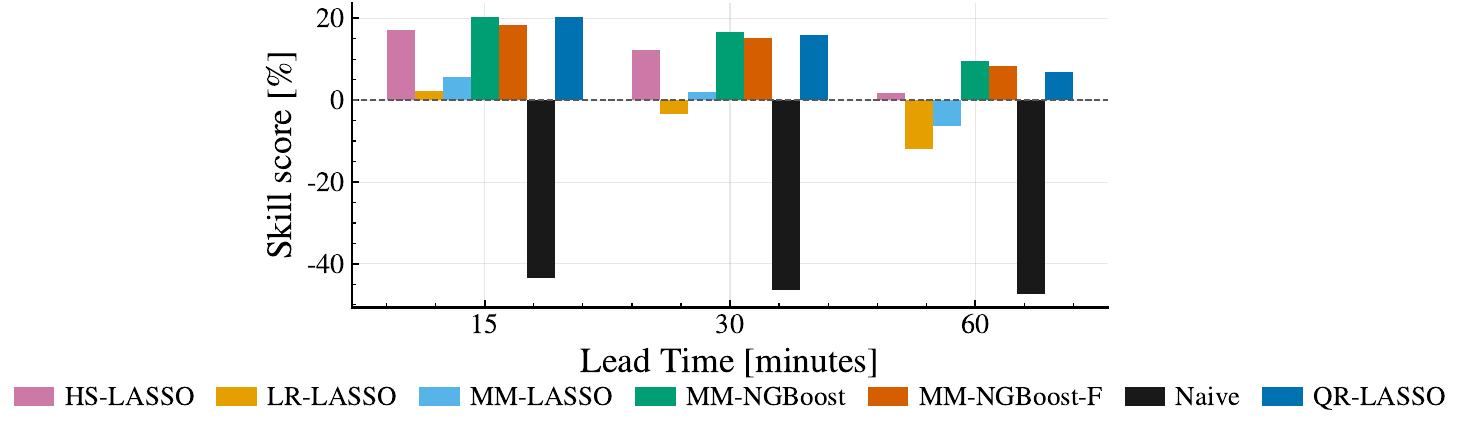}}
\caption{Mean Pinball Loss \eqref{eq: skill score} was used as a skill score to compare our tested models with the benchmark at each key lead time. The levels correspond to the batch forecasting setup. \red{With the exception of MM-NGBoost-F, which is trained on the same fundamentals-only feature set as QR-Benchmark, all models presented here utilise information from the intraday market, whereas the reference method, QR-Benchmark, does not.} 
The skill score for QR-LASSO represents the gain from incorporating features derived from intraday trades as QR-LASSO and QR-Benchmark differ only in the inclusion of these features. \red{Similarly, the gap between MM-NGBoost-F and MM-NGBoost isolates this same effect within the mixture-model paradigm.}
\label{fig: skill score}}
\end{figure}

 The relative performance of all models is summarised in Figure~\ref{fig: skill score}, which reports the Mean Pinball Loss Skill Score relative to the QR benchmark. The results highlight the importance of accounting for the heteroskedastic nature of the price spread, as well as the substantial performance gains achievable by incorporating information from intraday trading. Beyond forecasting accuracy, the proposed MM-NGBoost framework also offers notable computational advantages: it is a parsimonious approach requiring only three models (one for system state classification and two for modelling the price spread under long and short conditions). In contrast, QR-LASSO requires estimating a separate model for each quantile; in our implementation, 19 quantiles were considered, resulting in training and tuning times approximately 5--10 times longer than those of MM-NGBoost. All models were trained on university high-performance computing clusters (e.g., Ubuntu Server~22.04 running on Dell Precision Rack~7920 nodes equipped with Intel Xeon Gold~6240R CPUs with 24 cores and 128\,GB RAM).

\section{Conclusion}
\label{sec: CONCLUSION}

\par We have explored the challenges of predicting the electricity imbalance price in Great Britain's single price balancing market during the expansion of weather-dependent renewables, which introduce significant uncertainty into short-term electricity markets. By incorporating data from continuous intraday trading into probabilistic forecasting frameworks, we have shown that it is possible to improve forecasts of the imbalance price by 20\% 30 seconds before gate-closure to 7.5\% one hour ahead of delivery. Furthermore, our analysis shows that the benefits are two-fold: improving the precision of forecasts, and improving the discrimination between periods of high and low forecast uncertainty, both of which are valuable for energy trading.

\par Incorporating features derived from intraday trade data significantly enhances the performance of state-of-the-art quantile regression models. Furthermore, we have proposed a mixture model that predicts the both the mean and variance of the price spread conditional on price regime and probability of the system being long or short. The approach performs comparably to state-of-the-art quantile regression models overall with slight advantages at longer lead-times while also offering substantial computational benefits.

\par Moreover, it is important to note that our case study period includes the COVID-19 pandemic and the Russia-Ukraine war, which induced volatile periods in electricity markets from late 2021 to early 2023. This adds another layer of complexity and highlights the resilience of our forecasting models during challenging times.

\par While, to the best of our knowledge, this is the first academic paper demonstrating the benefit of intraday trade data for imbalance price forecasting, it is likely that practitioners already exploit this result, particularly in the context of algorithmic trading. \red{The cost of purchasing intraday market data is substantial, even for research purposes.} This is perhaps one reason for the relatively small literature on intraday price forecasting. As well as slowing innovation, it prevents the best practice of enabling fully reproducible research.

\section*{Software}
\label{sec: software}
All models and empirical analyses in this study were implemented in Python (version 3.11). \red{To support reproducibility, we report the versions of the main libraries used for data processing, model estimation and evaluation.} Core numerical computations were carried out using NumPy \red{(1.26.4)} and SciPy \red{(1.13.0)}, with data handling in pandas \red{(2.2.2)}. \red{scikit-learn (1.4.2) was used for baseline model implementations, preprocessing and evaluation utilities, and statsmodels (0.14.2) for the quantile regression estimators. The Natural Gradient Boosting models were implemented using the NGBoost library (0.5.1), and probabilistic regression interfaces were provided by skpro (2.2.2). Hyperparameter optimization was performed using scikit-optimize (0.10.1) and Optuna (4.0.0), and feature-attribution analysis used SHAP (0.49.1).} All figures and visualizations were generated with Matplotlib \red{(3.8.4)} and seaborn \red{(0.13.2)}. \red{Dependencies were managed with Poetry to ensure a reproducible environment.}

\section*{Acknowledgements}
\label{sec: acknowledgements}

Data availability: Underlying data are available as follows: BMRS data \copyright~Elexon Limited copyright and database right [2024]\hyperlink{https://www.bmreports.com/}{https://www.bmreports.com/} subject to terms and conditions. Day-ahead and intraday market data were purchased from EPEX
SPOT \hyperlink{https://webshop.eex-group.com/}{https://webshop.eex-group.com/}, accessed 24 January 2024, and cannot be shared due to terms and conditions of access.
Due to confidentiality agreements with industrial partners, the code used to produce the results presented here cannot be shared.

For the purpose of open access, the authors have applied a Creative Commons Attribution (CC BY) licence to any Author Accepted Manuscript version arising from this submission.



\bibliography{references}

@article{Fisher2019AllSimultaneously,
    title = {{All Models are Wrong, but Many are Useful: Learning a Variable's Importance by Studying an Entire Class of Prediction Models Simultaneously}},
    year = {2019},
    journal = {Journal of Machine Learning Research},
    author = {Fisher, Aaron and Rudin, Cynthia and Dominici, Francesca},
    number = {177},
    pages = {1--81},
    volume = {20},
    doi = {http://jmlr.org/papers/v20/18-760.html}
}

@article{Zhao2020AnGeneration,
    title = {{An Adaptive Bilevel Programming Model for Nonparametric Prediction Intervals of Wind Power Generation}},
    year = {2020},
    journal = {IEEE Transactions on Power Systems},
    author = {Zhao, Changfei and Wan, Can and Song, Yonghua},
    number = {1},
    month = {1},
    pages = {424--439},
    volume = {35},
    doi = {10.1109/TPWRS.2019.2924355},
    issn = {0885-8950}
}

@inproceedings{Shinde2021AnalysingMarkets,
    title = {{Analysing trading trends in continuous intraday electricity markets}},
    year = {2021},
    booktitle = {2021 56th International Universities Power Engineering Conference: Powering Net Zero Emissions, UPEC 2021 - Proceedings},
    author = {Shinde, Priyanka and Kouveliotis-Lysikatos, Iasonas and Amelin, Mikael},
    month = {8},
    publisher = {Institute of Electrical and Electronics Engineers Inc.},
    isbn = {9781665443890},
    doi = {10.1109/UPEC50034.2021.9548168}
}

@article{Bunn2021AnalysisMarket,
    title = {{Analysis of the Fundamental Predictability of Prices in the British Balancing Market}},
    year = {2021},
    journal = {IEEE Transactions on Power Systems},
    author = {Bunn, Derek W. and Inekwe, John N. and Macgeehan, David},
    number = {2},
    month = {3},
    pages = {1309--1316},
    volume = {36},
    publisher = {Institute of Electrical and Electronics Engineers Inc.},
    doi = {10.1109/TPWRS.2020.3015871},
    issn = {15580679}
}

@article{Filardo1998BusinessDurations,
    title = {{Business cycle durations}},
    year = {1998},
    journal = {Journal of Econometrics},
    author = {Filardo, Andrew J. and Gordon, Stephen F.},
    number = {1},
    month = {7},
    pages = {99--123},
    volume = {85},
    doi = {10.1016/S0304-4076(97)00096-1},
    issn = {03044076}
}

@article{Diebold2002ComparingAccuracy,
    title = {{Comparing Predictive Accuracy}},
    year = {2002},
    journal = {Journal of Business {\&} Economic Statistics},
    author = {Diebold, Francis X and Mariano, Robert S},
    number = {1},
    month = {1},
    pages = {134--144},
    volume = {20},
    doi = {10.1198/073500102753410444},
    issn = {0735-0015}
}

@article{Machete2013ContrastingRules,
    title = {{Contrasting probabilistic scoring rules}},
    year = {2013},
    journal = {Journal of Statistical Planning and Inference},
    author = {Machete, Reason L.},
    number = {10},
    month = {10},
    pages = {1781--1790},
    volume = {143},
    doi = {10.1016/j.jspi.2013.05.012},
    issn = {03783758}
}

@article{Narajewski2020EconometricPrices,
    title = {{Econometric modelling and forecasting of intraday electricity prices}},
    year = {2020},
    journal = {Journal of Commodity Markets},
    author = {Narajewski, Michał and Ziel, Florian},
    month = {9},
    volume = {19},
    publisher = {Elsevier B.V.},
    doi = {10.1016/j.jcomm.2019.100107},
    issn = {24058513},
    arxivId = {1812.09081}
}

@article{Hong2020EnergyOutlook,
    title = {{Energy Forecasting: A Review and Outlook}},
    year = {2020},
    journal = {IEEE Open Access Journal of Power and Energy},
    author = {Hong, Tao and Pinson, Pierre and Wang, Yi and Weron, Rafal and Yang, Dazhi and Zareipour, Hamidreza},
    pages = {376--388},
    volume = {7},
    doi = {10.1109/OAJPE.2020.3029979},
    issn = {2687-7910}
}

@article{Messner2020EvaluationView,
    title = {{Evaluation of wind power forecasts—An up-to-date view}},
    year = {2020},
    journal = {Wind Energy},
    author = {Messner, Jakob W. and Pinson, Pierre and Browell, Jethro and Bjerreg{\aa}rd, Mathias B. and Schicker, Irene},
    number = {6},
    month = {6},
    pages = {1461--1481},
    volume = {23},
    publisher = {John Wiley and Sons Ltd},
    doi = {10.1002/we.2497},
    issn = {10991824}
}

@article{Janke2019ForecastingTrading,
    title = {{Forecasting the price distribution of continuous intraday electricity trading}},
    year = {2019},
    journal = {Energies},
    author = {Janke, Tim and Steinke, Florian},
    number = {22},
    month = {11},
    volume = {12},
    publisher = {MDPI AG},
    doi = {10.3390/en12224262},
    issn = {19961073}
}

@article{Groenwold2020InformativeKnowing,
    title = {{Informative missingness in electronic health record systems: the curse of knowing}},
    year = {2020},
    journal = {Diagnostic and Prognostic Research},
    author = {Groenwold, Rolf H. H.},
    number = {1},
    month = {12},
    pages = {8},
    volume = {4},
    doi = {10.1186/s41512-020-00077-0},
    issn = {2397-7523}
}

@inproceedings{Ke2017LightGBM:Tree,
    title = {{LightGBM: A Highly Efficient Gradient Boosting Decision Tree}},
    year = {2017},
    booktitle = {Advances in Neural Information Processing Systems},
    author = {Ke, Guolin and Meng, Qi and Finley, Thomas and Wang, Taifeng and Chen, Wei and Ma, Weidong and Ye, Qiwei and Liu, Tie-Yan},
    editor = {Guyon, I and Luxburg, U Von and Bengio, S and Wallach, H and Fergus, R and Vishwanathan, S and Garnett, R},
    pages = {},
    volume = {30},
    publisher = {Curran Associates, Inc.}
}

@inproceedings{Duan2020NGBoost:Prediction,
    title = {{NGBoost: Natural Gradient Boosting for Probabilistic Prediction}},
    year = {2020},
    booktitle = {Proceedings of the 37 th International Conference on Machine Learning},
    author = {Duan, Tony and Avati, Anand and Ding, Daisy Yi and Thai, Khanh K. and Basu, Sanjay and Ng, Andrew Y. and Schuler, Alejandro},
    month = {10},
    doi = {10.5555/3524938.3525190},
    arxivId = {1910.03225}
}

@article{Kullback1951OnSufficiency,
    title = {{On Information and Sufficiency}},
    year = {1951},
    journal = {The Annals of Mathematical Statistics},
    author = {Kullback, S. and Leibler, R. A.},
    number = {1},
    month = {3},
    pages = {79--86},
    volume = {22},
    doi = {10.1214/aoms/1177729694},
    issn = {0003-4851}
}

@article{Narajewski2021OptimalCosts,
    title = {{Optimal bidding in hourly and quarter-hourly electricity price auctions: trading large volumes of power with market impact and transaction costs}},
    year = {2021},
    journal = {Energy Economics},
    author = {Narajewski, Michał and Ziel, Florian},
    month = {4},
    doi = {doi.org/10.1016/j.eneco.2022.105974},
    arxivId = {2104.14204}
}

@inproceedings{Akiba2019Optuna,
    title = {{Optuna}},
    year = {2019},
    booktitle = {Proceedings of the 25th ACM SIGKDD International Conference on Knowledge Discovery {\&} Data Mining},
    author = {Akiba, Takuya and Sano, Shotaro and Yanase, Toshihiko and Ohta, Takeru and Koyama, Masanori},
    month = {7},
    pages = {2623--2631},
    publisher = {ACM},
    address = {New York, NY, USA},
    isbn = {9781450362016},
    doi = {10.1145/3292500.3330701}
}

@article{Browell2022PredictingOpportunities,
    title = {{Predicting Electricity Imbalance Prices and Volumes: Capabilities and Opportunities}},
    year = {2022},
    journal = {Energies},
    author = {Browell, Jethro and Gilbert, Ciaran},
    number = {10},
    month = {5},
    volume = {15},
    publisher = {MDPI},
    doi = {10.3390/en15103645},
    issn = {19961073}
}

@article{Lucas2020PriceRegression,
    title = {{Price forecasting for the balancing energy market using machine-learning regression}},
    year = {2020},
    journal = {Energies},
    author = {Lucas, Alexandre and Pegios, Konstantinos and Kotsakis, Evangelos and Clarke, Dan},
    number = {20},
    month = {10},
    volume = {13},
    publisher = {MDPI AG},
    doi = {10.3390/en13205420},
    issn = {19961073}
}

@article{Narajewski2022ProbabilisticPrices,
    title = {{Probabilistic Forecasting of German Electricity Imbalance Prices}},
    year = {2022},
    journal = {Energies},
    author = {Narajewski, Michał},
    number = {14},
    month = {7},
    volume = {15},
    publisher = {MDPI},
    doi = {10.3390/en15144976},
    issn = {19961073},
    arxivId = {2205.11439}
}

@article{Andrade2017ProbabilisticModel,
    title = {{Probabilistic price forecasting for day-ahead and intraday markets: Beyond the statistical model}},
    year = {2017},
    journal = {Sustainability},
    author = {Andrade, José R. and Filipe, Jorge and Reis, Marisa and Bessa, Ricardo J.},
    number = {11},
    month = {10},
    volume = {9},
    publisher = {MDPI},
    doi = {10.3390/su9111990},
    issn = {20711050}
}

@techreport{Pinson2006PropertiesEvaluation,
    title = {{Properties of Quantile and Interval Forecasts of Wind Generation and their Evaluation}},
    year = {2006},
    author = {Pinson, P and Kariniotakis, G and Nielsen, H Aa and Nielsen, T S and Madsen, H},
    institution = {Ecole des Mines de Paris}
}

@misc{Nowotarski2018RecentForecasting,
    title = {{Recent advances in electricity price forecasting: A review of probabilistic forecasting}},
    year = {2018},
    booktitle = {Renewable and Sustainable Energy Reviews},
    author = {Nowotarski, Jakub and Weron, Rafał},
    month = {1},
    pages = {1548--1568},
    volume = {81},
    publisher = {Elsevier Ltd},
    doi = {10.1016/j.rser.2017.05.234},
    issn = {18790690}
}

@article{Koenker1978RegressionQuantiles,
    title = {{Regression Quantiles}},
    year = {1978},
    journal = {Econometrica},
    author = {Koenker, Roger and Bassett, Gilbert},
    number = {1},
    month = {1},
    pages = {33},
    volume = {46},
    doi = {10.2307/1913643},
    issn = {00129682}
}

@article{Uniejewski2021RegularizedForecasting,
    title = {{Regularized quantile regression averaging for probabilistic electricity price forecasting}},
    year = {2021},
    journal = {Energy Economics},
    author = {Uniejewski, Bartosz and Weron, Rafał},
    month = {3},
    pages = {105121},
    volume = {95},
    publisher = {North-Holland},
    doi = {10.1016/J.ENECO.2021.105121},
    issn = {0140-9883}
}

@article{Browell2018RiskMarket,
    title = {{Risk Constrained Trading Strategies for Stochastic Generation with a Single-Price Balancing Market}},
    year = {2018},
    journal = {Energies},
    author = {Browell, Jethro},
    number = {6},
    month = {5},
    pages = {1345},
    volume = {11},
    doi = {10.3390/en11061345},
    issn = {1996-1073}
}

@misc{Head2022Scikit-optimize/scikit-optimize,
    title = {{scikit-optimize/scikit-optimize}},
    year = {2022},
    author = {Head, Tim and Kumar, Manoj and Nahrstaedt, Holger and Louppe, Gilles and Shcherbatyi, Iaroslav},
    month = {4},
    publisher = {Zenodo},
    doi = {10.5281/zenodo.6451894}
}

@article{Gneiting2007StrictlyEstimation,
    title = {{Strictly proper scoring rules, prediction, and estimation}},
    year = {2007},
    journal = {Journal of the American Statistical Association},
    author = {Gneiting, Tilmann and Raftery, Adrian E.},
    number = {477},
    month = {3},
    pages = {359--378},
    volume = {102},
    doi = {10.1198/016214506000001437},
    issn = {01621459}
}

@article{Goodarzi2019ThePrices,
    title = {{The impact of renewable energy forecast errors on imbalance volumes and electricity spot prices}},
    year = {2019},
    journal = {Energy Policy},
    author = {Goodarzi, Shadi and Perera, H. Niles and Bunn, Derek},
    month = {11},
    pages = {110827},
    volume = {134},
    publisher = {Elsevier},
    doi = {10.1016/J.ENPOL.2019.06.035},
    issn = {0301-4215}
}

@article{Pinson2007TradingPower,
    title = {{Trading Wind Generation From Short-Term Probabilistic Forecasts of Wind Power}},
    year = {2007},
    journal = {IEEE Transactions on Power Systems},
    author = {Pinson, Pierre and Chevallier, Christophe and Kariniotakis, George N.},
    number = {3},
    month = {8},
    pages = {1148--1156},
    volume = {22},
    doi = {10.1109/TPWRS.2007.901117},
    issn = {0885-8950}
}

@article{Bottieau2020Very-Short-TermSettlement,
    title = {{Very-Short-Term Probabilistic Forecasting for a Risk-Aware Participation in the Single Price Imbalance Settlement}},
    year = {2020},
    journal = {IEEE Transactions on Power Systems},
    author = {Bottieau, Jeremie and Hubert, Louis and De Greve, Zacharie and Vallee, Francois and Toubeau, Jean-Francois},
    number = {2},
    month = {3},
    pages = {1218--1230},
    volume = {35},
    doi = {10.1109/TPWRS.2019.2940756},
    issn = {0885-8950}
}
\newpage
\section*{A Permutation Feature Importance}
\label{sec: PFI}

\begin{figure}[htbp]
    \centering
    \includegraphics[width=0.48\textwidth]{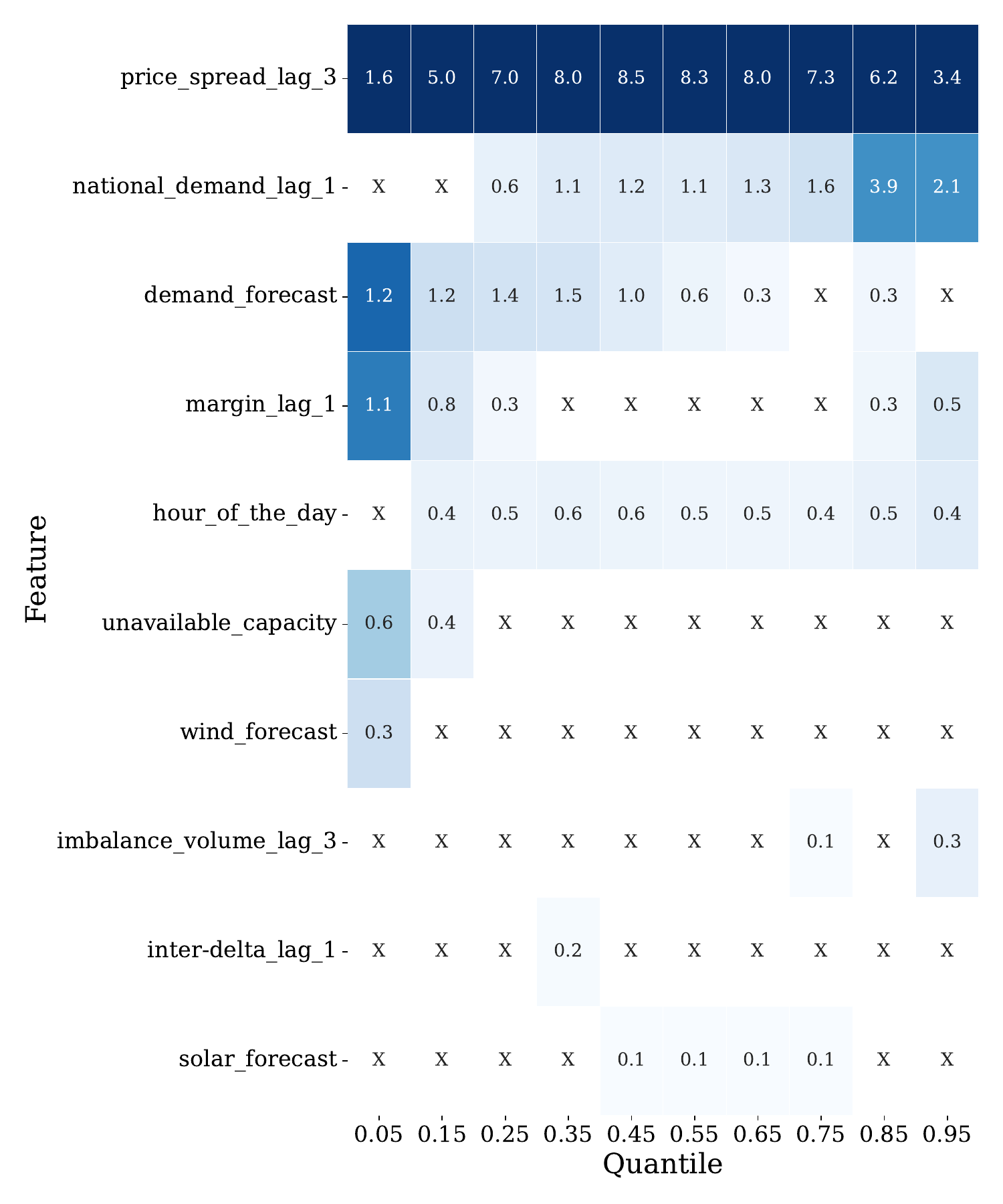}\hfill
    \includegraphics[width=0.48\textwidth]{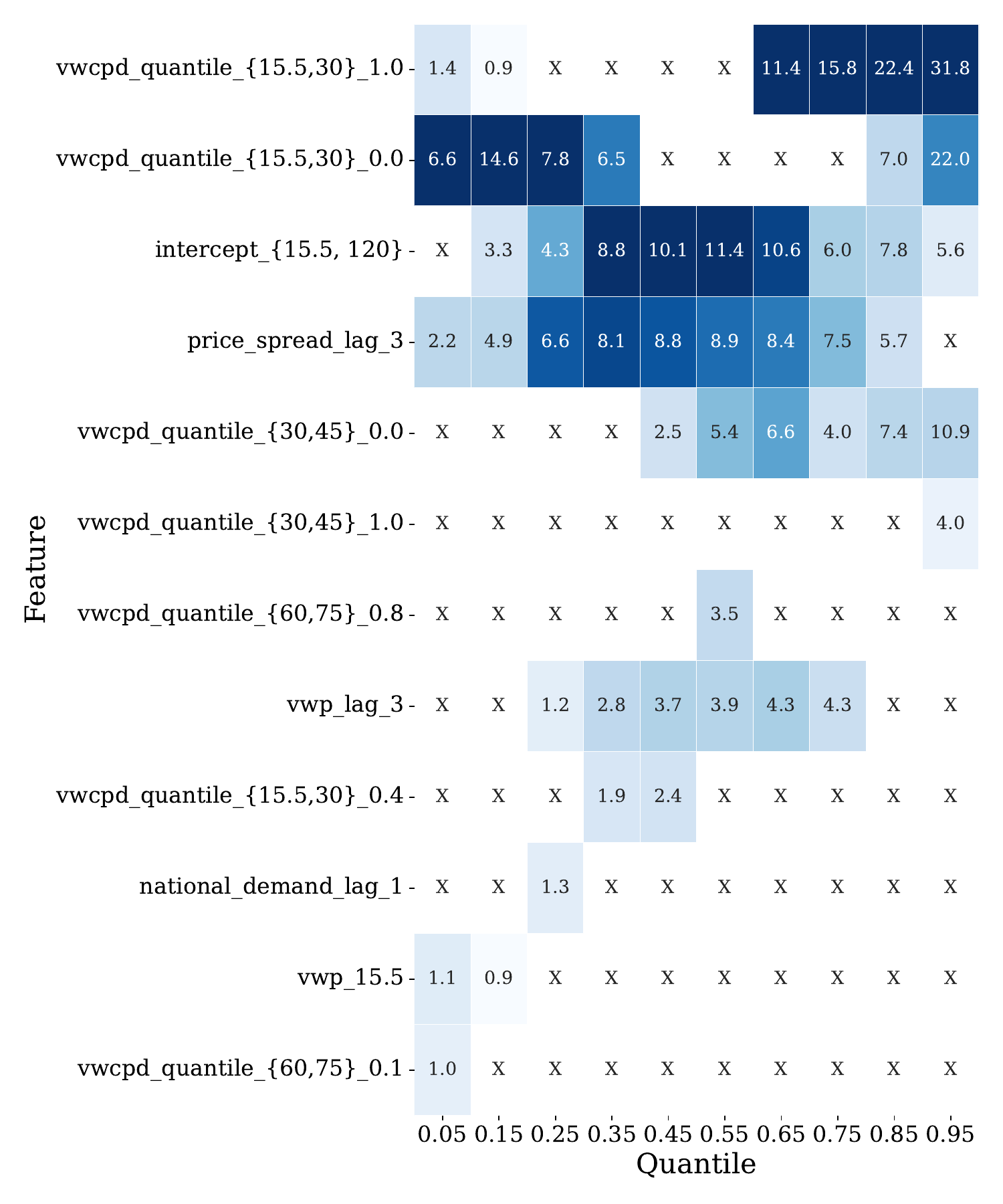}
    \caption{Permutation feature importance (PFI) heatmaps for a lead time of 15.5 minutes. The left panel corresponds to QR--Benchmark, which relies exclusively on features constructed from publicly available system and market variables and does not include covariates derived from intraday market data. In contrast, the right panel reports results for the QR--LASSO model augmented with intraday market-based features. While both models identify lagged price spread and demand-related variables as relevant predictors across quantiles, the inclusion of intraday-derived features in the right panel leads to a richer and more heterogeneous importance structure, particularly at higher quantile levels.}
    \label{fig:PFI 1}
\end{figure}

\begin{figure}[htbp]
    \centering
    \includegraphics[width=0.48\textwidth]{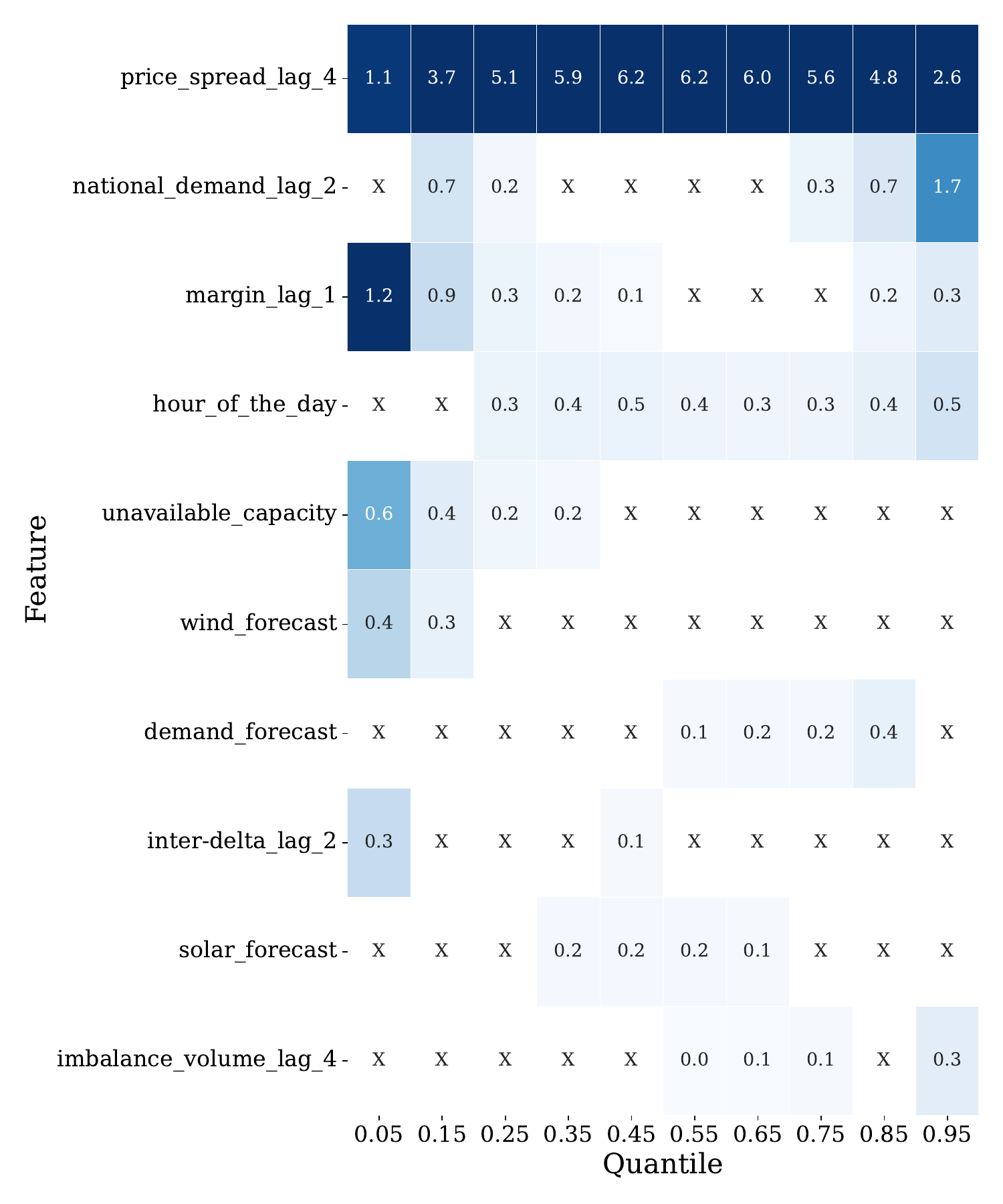}\hfill
    \includegraphics[width=0.48\textwidth]{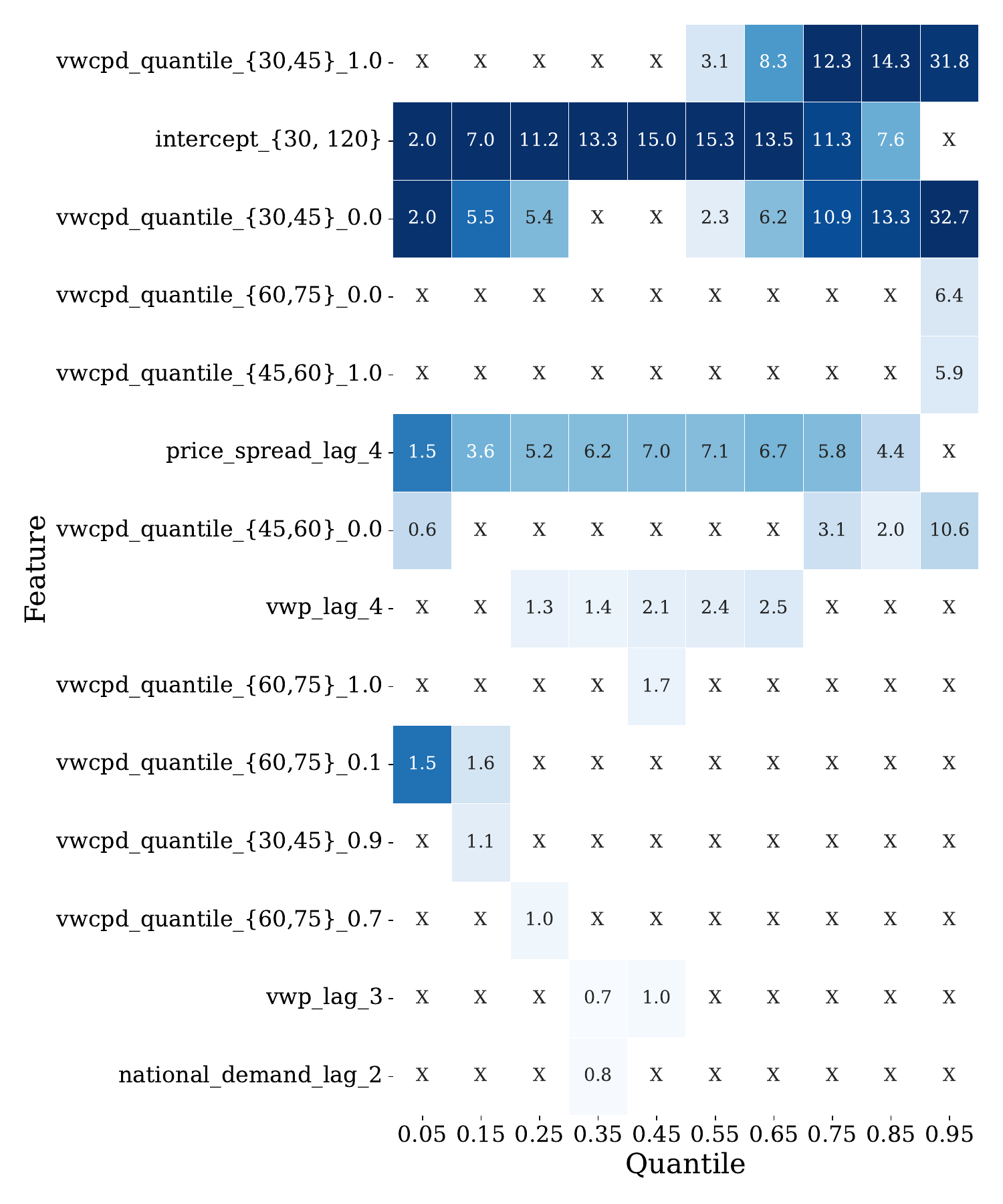}
    \caption{Permutation feature importance (PFI) heatmaps for a 30-minute lead time. The left panel reports results for QR--Benchmark, which excludes features derived from intraday market data, while the right panel corresponds to the QR--LASSO model augmented with intraday market-based covariates. At this longer forecasting horizon, lagged price spread and demand-related variables remain important predictors across quantiles in both models. The inclusion of intraday-derived features in the right panel continues to yield a more diverse importance pattern, with several market-based quantile features contributing to forecast performance, particularly in the upper tail of the conditional distribution.}
    \label{fig:PFI 2}
\end{figure}

\begin{figure}[htbp]
    \centering
    \includegraphics[width=0.48\textwidth]{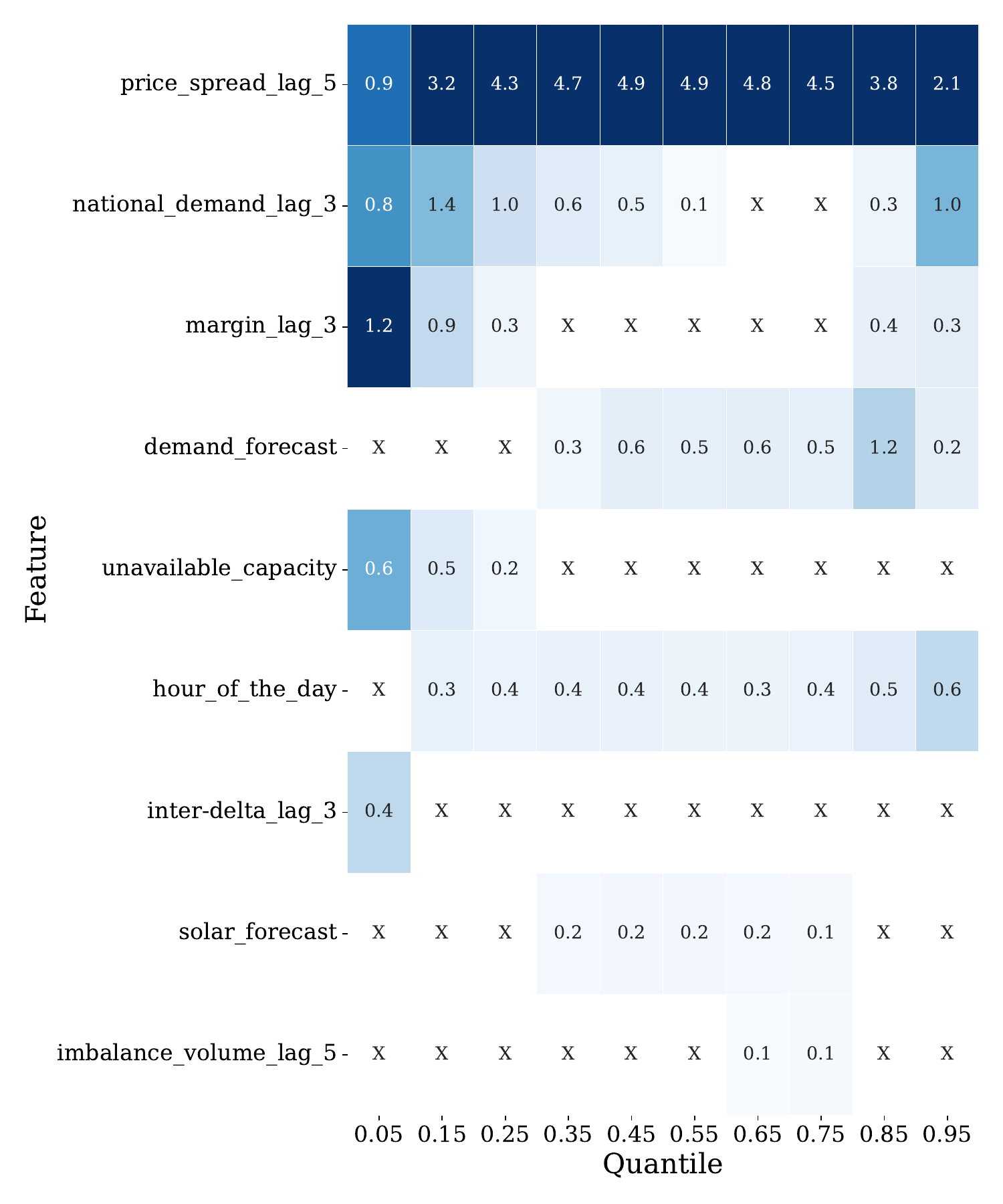}\hfill
    \includegraphics[width=0.48\textwidth]{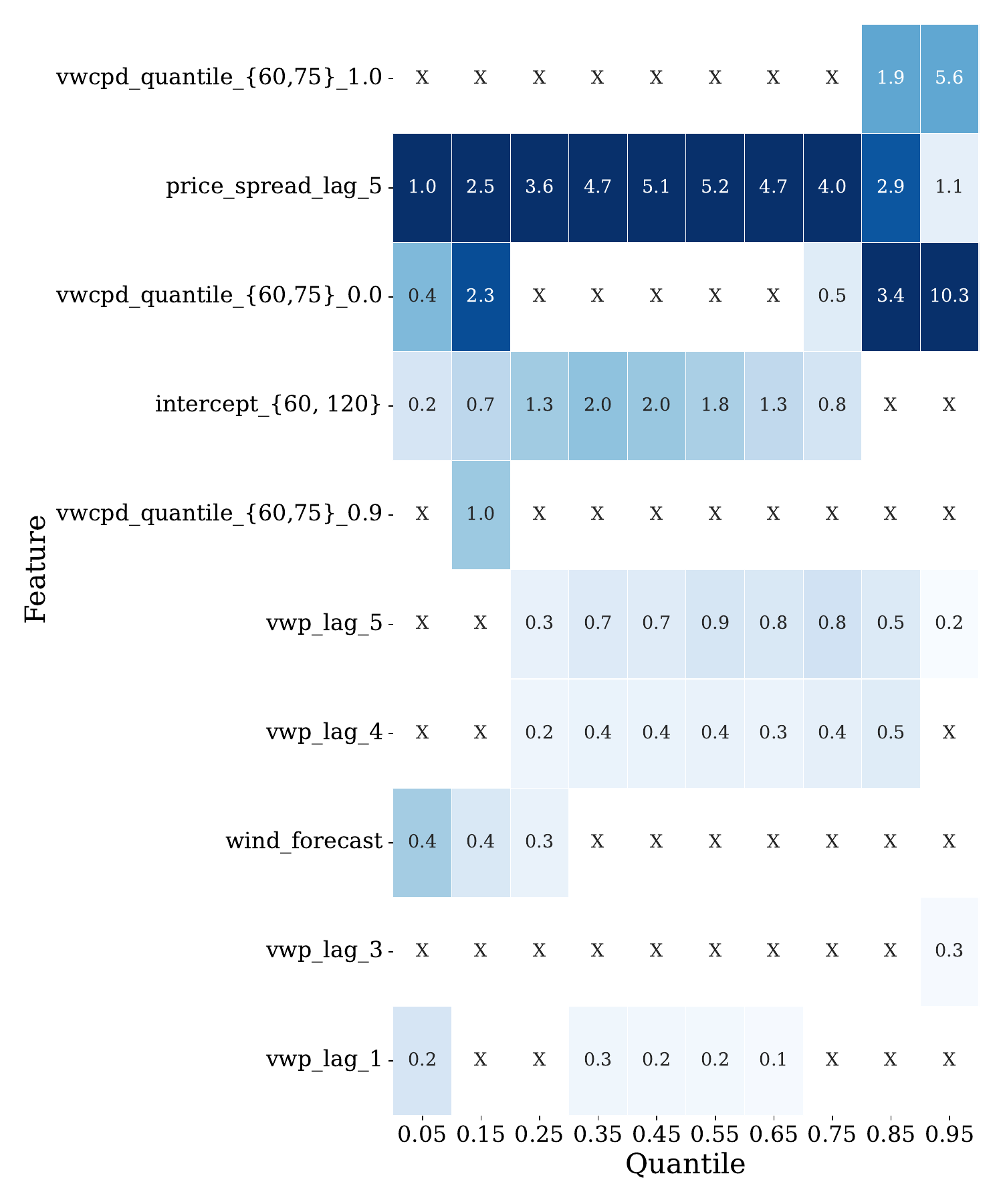}
    \caption{Permutation feature importance (PFI) heatmaps for a 60-minute lead time. The left panel shows results for QR--Benchmark, which excludes intraday market-derived features, while the right panel corresponds to the QR--LASSO model augmented with intraday market-based covariates. At this longer forecasting horizon, the importance of lagged price spread and system-level variables remains evident in both models. However, the QR--LASSO model exhibits a more pronounced contribution from intraday market-based quantile features, particularly at higher quantiles, indicating their relevance for characterizing tail behavior at extended lead times.}
    \label{fig:PFI}
\end{figure}

Figures~\ref{fig:PFI 1}--\ref{fig:PFI} present the permutation feature importance (PFI) \citep{Fisher2019AllSimultaneously} analysis conducted on the test set, with the aim of comparing the relative contribution of input features in QR-Benchmark and QR-LASSO. The latter is selected as the proposed method, as it consistently ranks among the top-performing models in terms of predictive accuracy. PFI quantifies feature relevance by measuring the degradation in pinball loss when the values of a given feature are randomly permuted, thereby breaking its association with the target variable. Since QR-LASSO exhibits stronger and more stable predictive performance, the corresponding PFI results are expected to provide a more reliable characterization of feature relevance. The analysis is performed for our selected lead times in order to assess how feature importance evolves as the forecasting horizon increases.

\par \red{Concretely, for a given quantile level $\tau$ we first evaluate the pinball loss of the trained model on the unperturbed test set. For each feature in turn, we then randomly permute that feature's values across the test observations--leaving all other features unchanged--and re-evaluate the pinball loss; this is repeated over 100 random permutations and the results averaged. The permutation importance of the feature at quantile $\tau$ is the resulting mean increase in pinball loss relative to the unperturbed baseline: a large increase indicates that the model relies heavily on that feature to produce accurate forecasts at that quantile, whereas an increase close to zero indicates that the feature carries little quantile-specific information. Each cell of the heatmaps is annotated with this raw permutation importance (the mean increase in pinball loss) for one feature (rows) at one quantile level (columns). The cell shading, by contrast, encodes the importance after scaling within each quantile by that quantile's maximum importance, so that the colour is comparable across quantile columns whose baseline losses differ; darker cells thus correspond to the features that the model relies on most at the corresponding quantile. Cells marked ``X'' denote feature--quantile combinations for which the feature does not appear among the most important features at that quantile (see below).}

\par \red{Because the full feature set is large and most features are uninformative at any given quantile, displaying every feature would obscure the relevant structure. For each model and lead time we therefore retain, at each quantile, the five features with the highest permutation importance at that quantile, and the heatmap rows are the union of these per-quantile sets across all quantiles. The rows are ordered by each feature's mean importance across quantiles, with the most important on average at the top. A feature that is among the top five at some quantiles but not at others is shown with an ``X'' in the columns where it does not appear, which is why the heatmaps are sparse. This selection is performed independently for each model, which is why the feature sets differ between the QR-Benchmark and QR-LASSO heatmaps and across lead times; these differences are themselves informative, as they reveal which features each model comes to rely on. The row ordering serves only to aid visual interpretation and does not affect the computed importances.}

Feature labels in the right panel follow the notation \text{feature}\_{\{a,b\}\_q}
, where $a$ and $b$ denote the start and end (in minutes) of the aggregation window prior to the forecast origin, and $q$ denotes the quantile level. For instance, \text{vwcpd\_quantile}\_{\{15.5,30\}\_{1.0}} represents the volume-weighted cumulative probability distribution quantile at level $1.0$ computed over the interval $[15.5,30]$ minutes before the target period.

\red{In Figures~\ref{fig:PFI 1}--\ref{fig:PFI}, the y-axis denotes the input features and the x-axis the quantile level. Each cell is annotated with the raw permutation importance and shaded by its per-quantile scaled value: darker cells indicate stronger degradation in pinball loss when the feature is permuted, and cells marked ``X'' indicate features not ranked among the most important at that quantile.}

\par \red{The heatmaps reveal a clear difference in how the two models form their forecasts. For the QR-Benchmark, which has no access to intraday trade data, importance is concentrated almost entirely on the lagged price spread (\text{price\_spread\_lag\_3}), supported by demand-side variables such as \text{national\_demand\_lag\_1} and \text{demand\_forecast}; the model essentially extrapolates from the most recent realised spread and the anticipated system demand. Once intraday trade features are made available in QR-LASSO, the picture changes markedly. The features derived from the volume-weighted price distribution of recent trades become the dominant predictors, and they do so primarily in the tails of the forecast distribution: the upper edge of the traded-price distribution in the final window before gate closure (\text{vwcpd\_quantile}$_{\{15.5,30\}\_1.0}$) is by far the most important feature at the highest quantiles, while the corresponding lower edge (\text{vwcpd\_quantile}$_{\{15.5,30\}\_0.0}$) dominates at the lowest quantiles. The VWAP-trend feature (\text{intercept}$_{\{15.5,120\}}$) and the lagged price spread retain importance around the centre of the distribution. In other words, the most recent trades are especially informative about the extreme quantiles of the imbalance price, which is precisely where the largest imbalance-price risk lies. This helps explain both why the intraday features deliver their largest improvements at the shortest lead time and why the gains are most pronounced in the tails of the predictive distribution, consistent with the pinball-loss results reported in Section~\ref{sec: model performance}.}

\newpage
\section*{B Diebold-Mariano Test Results}
\label{sec: DM test}

\begin{figure}[htbp]
    \centering

    \begin{subfigure}[t]{0.45\textwidth}
        \centering
        \includegraphics[width=\linewidth]{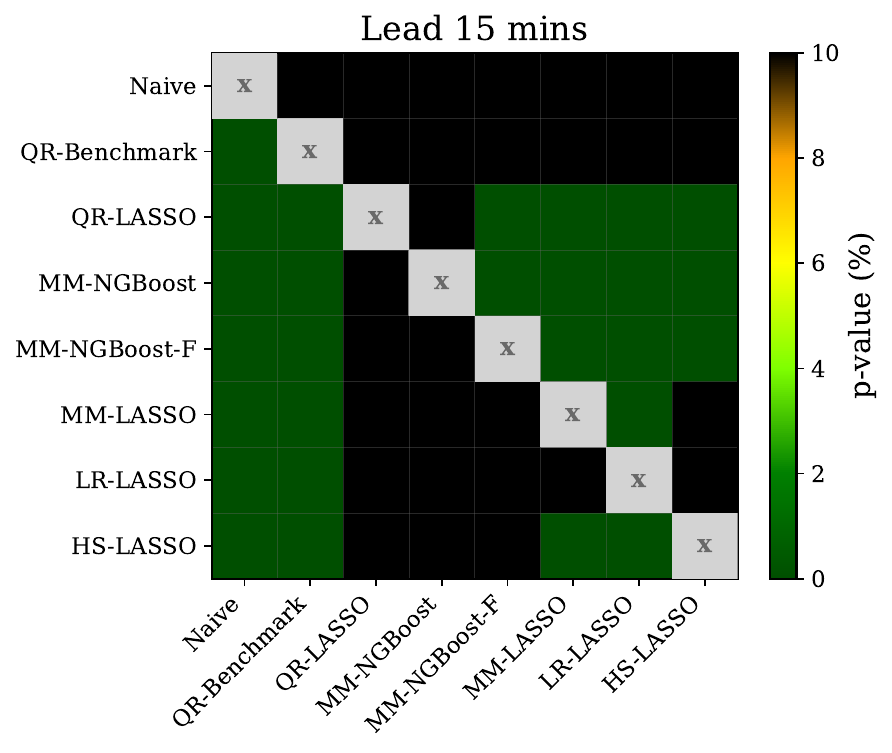}
        \caption{Lead time 15.5 mins}
        \label{fig:DM 15}
    \end{subfigure}
    \hfill
    \begin{subfigure}[t]{0.45\textwidth}
        \centering
        \includegraphics[width=\linewidth]{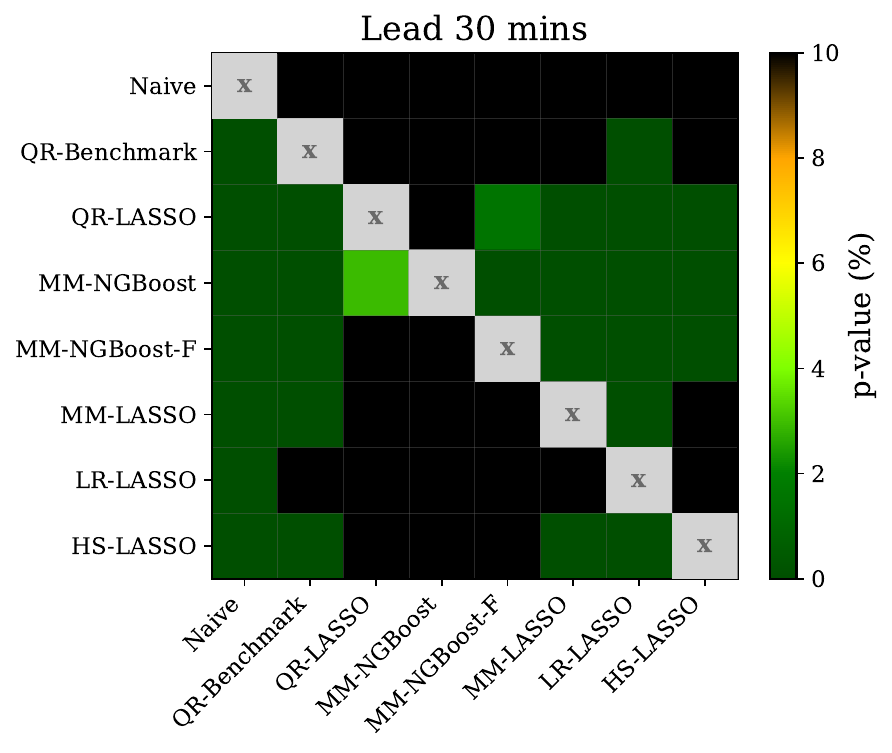}
        \caption{Lead time 30 mins}
        \label{fig: DM 30}
    \end{subfigure}

    \vspace{1em}

    \begin{subfigure}[t]{0.45\textwidth}
        \centering
        \includegraphics[width=\linewidth]{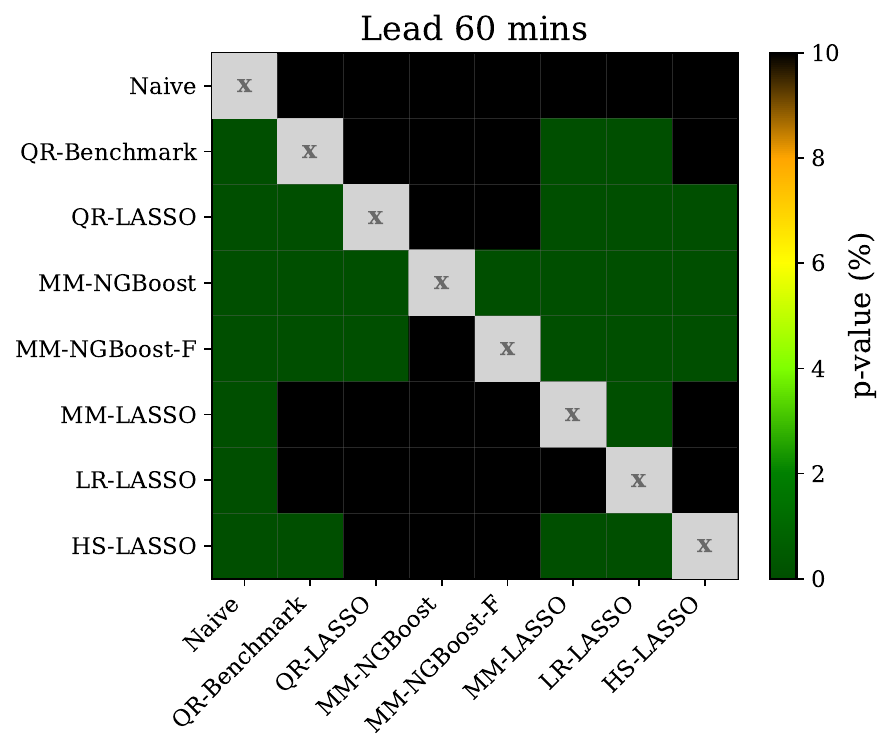}
        \caption{Lead time 60 mins}
        \label{fig:DM 60}
    \end{subfigure}

    \caption{Diebold--Mariano test results across all methods and lead times. \red{MM-NGBoost-F, the fundamentals-only variant of MM-NGBoost, is included alongside the other models.} Figure displays a chessboard-shaded heatmap of p-values, indicating whether the forecasts of the methods shown on the y-axis are significantly more accurate than those shown on the x-axis. \red{P-values close to zero indicate that the forecasts of the method on the y-axis are significantly more accurate than those of the method on the x-axis. The proposed models, QR-LASSO and MM-NGBoost, produce significantly better forecasts than the QR-Benchmark across all lead times.}}
    \label{fig: Diebold-Mariano test}
\end{figure}

In Figure \ref{fig: Diebold-Mariano test}, we present additional results from the Diebold--Mariano test \citep{Diebold2002ComparingAccuracy}. The predictive accuracy of each method is compared across all lead times. Figure \ref{fig: Diebold-Mariano test} displays a chessboard-shaded heatmap of p-values, indicating whether the forecasts of the methods shown on the y-axis are significantly more accurate than those shown on the x-axis. P-values close to zero indicate that the forecasts of the method on the y-axis are significantly more accurate than those of the method on the x-axis.

At the 15-minute lead time, the Diebold--Mariano test results point to the presence of statistically significant differences among several forecasting approaches. Both QR-LASSO and MM-NGBoost deliver significantly more accurate forecasts than the QR-Benchmark, while MM-LASSO also demonstrates improved predictive performance relative to the benchmark specification. By contrast, performance differences among the mixture-based models are largely not statistically significant\red{, with the notable exception of MM-NGBoost against MM-NGBoost-F, discussed below,} and the parametric LR-LASSO model fails to exhibit any systematic advantage.

At the 30-minute lead time, the Diebold--Mariano test results suggest a more distinct ordering of the competing models. QR-LASSO and MM-NGBoost continue to outperform the QR-Benchmark and the remaining mixture-based and parametric LASSO variants with strong statistical support. Outside these leading methods, differences across the remaining approaches are mostly insignificant, indicating broadly comparable levels of predictive accuracy.

At the 60-minute lead time, the Diebold--Mariano test results display the most pronounced separation among forecasting methods. QR-LASSO and MM-NGBoost emerge as the clearly dominant specifications, significantly outperforming the QR-Benchmark as well as all alternative model formulations. Also, we distinguish the MM-NGBoost as the dominant method across all methods. Additionally, the LR-LASSO model is consistently outperformed. Taken together, these results highlight that the relative performance gains of MM-NGBoost become increasingly substantial as the forecast horizon lengthens.

\par \red{We additionally use the Diebold--Mariano test to assess the significance of the trade-data effect within the mixture-model paradigm, by comparing MM-NGBoost against MM-NGBoost-F (Section~\ref{sec: model performance}). MM-NGBoost produces significantly more accurate forecasts than MM-NGBoost-F at every lead time under both the batch and fixed training set-ups, confirming that the improvement from including intraday trade features, while modest in magnitude, is statistically significant throughout.}

\par \red{We similarly use the Diebold--Mariano test to compare HS-LASSO against LR-LASSO and against QR-Benchmark (Section~\ref{subsec: HS-LASSO}). HS-LASSO produces significantly more accurate forecasts than both LR-LASSO and QR-Benchmark at every lead time and under both training set-ups, while the reverse comparisons are not significant in either case. This confirms that replacing the Gaussian assumption in LR-LASSO with a historical-simulation approach yields a statistically significant improvement using an identical point forecast, and that HS-LASSO constitutes a benchmark that significantly outperforms QR-Benchmark without requiring intraday trade data.}

\newpage

\section*{C Point Forecast Evaluation Metrics}
\label{sec: Evaluation Metrics}

Point forecast accuracy is assessed using the mean absolute error (MAE) and the root mean squared error (RMSE), which capture complementary aspects of forecast performance. For all probabilistic models considered, point forecasts are obtained from the median (0.5-quantile) of the predictive distribution.

Tables~\ref{tab:mae_rmse_static} and~\ref{tab:mae_rmse_batch} summarise MAE and RMSE values for lead times of 15, 30, and 60 minutes. Across both tables, methods based on distributional modelling generally outperform benchmark and linear alternatives, particularly at shorter and medium lead times. In several cases, improvements are observed simultaneously in terms of MAE and RMSE, indicating gains in both average accuracy and robustness to larger forecast errors.

The RMSE of the LR-LASSO model is quite close or even lower for lead time 60 for the batch setup, compared with the 0.5-quantile models, which is an expected outcome given the objectives of the respective estimation procedures. RMSE is a metric aligned with the conditional mean and is therefore naturally associated with expectation-based methods such as OLS. LR-LASSO is the only model in this comparison that is explicitly optimised for this objective. In contrast, the remaining models are estimated by minimising the pinball loss at the 0.5 quantile, which is equivalent to minimising the MAE.

Overall, the results suggest that exploiting distributional information in the forecasting stage leads to systematically more accurate point predictions across lead times. Performance differences between methods become more pronounced as the forecasting horizon increases, underscoring the importance of model choice when predicting imbalance prices further ahead. \red{MM-NGBoost-F, the fundamentals-only variant of MM-NGBoost (Section~\ref{sec: model performance}), achieves point-forecast accuracy close to MM-NGBoost and QR-LASSO despite lacking intraday trade features, consistent with the modest marginal contribution of these features observed for the probabilistic metrics. HS-LASSO (Section~\ref{subsec: HS-LASSO}) achieves MAE and RMSE values close to those of LR-LASSO at every lead time, which is expected since both models share the same underlying point forecast and differ only in how quantiles are constructed around it. This confirms that the substantial difference in probabilistic performance between HS-LASSO and LR-LASSO documented in Section~\ref{sec: model performance} arises entirely from the treatment of the residual distribution, and not from any difference in point-forecast accuracy.}

\begin{table}[htbp]
\centering
\caption{MAE and RMSE across lead times under the static forecasting framework.
For each method, point forecasts are obtained from the median (0.5-quantile) of the predictive distribution and evaluated against the realized observations at each lead time. The best-performing method in each column is shown in bold. \red{MM-NGBoost-F, the fundamentals-only variant of MM-NGBoost, and HS-LASSO, which shares its point forecast with LR-LASSO but constructs quantiles via historical simulation, are included alongside the other models.}}
\label{tab:mae_rmse_static}
\begin{tabular}{lrrrrrr}
\toprule
 & \multicolumn{3}{c}{MAE} & \multicolumn{3}{c}{RMSE} \\
Lead Time & 15 & 30 & 60 & 15 & 30 & 60 \\
\midrule
\red{Naive} & \red{41.5} & \red{45.1} & \red{47.6} & \red{63.2} & \red{66.5} & \red{69.2} \\
QR-Benchmark & 37.8 & 39.9 & 41.0 & 55.4 & 56.5 & 57.2 \\
LR-LASSO & 33.1 & 36.1 & 39.3 & 45.1 & 49.2 & \textbf{53.0} \\
\red{HS-LASSO} & \red{32.8} & \red{36.0} & \red{39.5} & \red{44.8} & \red{49.1} & \red{53.5} \\
MM-LASSO & 34.8 & 37.9 & 41.2 & 46.8 & 50.4 & 55.6 \\
MM-NGBoost & \textbf{31.0} & 34.7 & 39.2 & 45.0 & 49.3 & 53.9 \\
\red{MM-NGBoost-F} & \red{31.8} & \red{35.4} & \red{39.7} & \red{46.2} & \red{50.2} & \red{54.2} \\
QR-LASSO & 31.3 & \textbf{34.4} & \textbf{39.0} & \textbf{44.3} & \textbf{47.8} & 53.7 \\
\bottomrule
\end{tabular}
\end{table}

\begin{table}[htbp]
\centering
\caption{MAE and RMSE across lead times under the batch forecasting framework.
For each method, point forecasts are obtained from the median (0.5-quantile) of the predictive distribution and evaluated against the realized observations at each lead time. The best-performing method in each column is shown in bold. \red{MM-NGBoost-F, the fundamentals-only variant of MM-NGBoost, and HS-LASSO, which shares its point forecast with LR-LASSO but constructs quantiles via historical simulation, are included alongside the other models.}}
\label{tab:mae_rmse_batch}
\begin{tabular}{lrrrrrr}
\toprule
 & \multicolumn{3}{c}{MAE} & \multicolumn{3}{c}{RMSE} \\
Lead Time & 15 & 30 & 60 & 15 & 30 & 60 \\
\midrule
\red{Naive} & \red{41.5} & \red{45.1} & \red{47.6} & \red{63.2} & \red{66.5} & \red{69.2} \\
QR-Benchmark & 37.8 & 39.9 & 40.9 & 55.4 & 56.4 & 57.1 \\
LR-LASSO & 32.2 & 35.6 & 39.8 & 44.6 & 49.1 & 53.9 \\
\red{HS-LASSO} & \red{32.0} & \red{35.5} & \red{39.9} & \red{44.5} & \red{49.1} & \red{54.2} \\
MM-LASSO & 33.8 & 37.0 & 40.9 & 45.8 & 49.4 & 55.2 \\
MM-NGBoost & \textbf{31.0} & 34.4 & 39.0 & 45.0 & 49.1 & 53.7 \\
\red{MM-NGBoost-F} & \red{31.3} & \red{34.8} & \red{39.2} & \red{45.9} & \red{49.8} & \red{54.0} \\
QR-LASSO & 31.1 & \textbf{34.2} & \textbf{38.9} & \textbf{44.0} & \textbf{47.7} & \textbf{53.6} \\
\bottomrule
\end{tabular}
\end{table}

\section*{D Fixed Training Results}

In this section we report the forecast performance under a fixed training set-up and contrasts it with the monthly batch training approach used in the main analysis. Overall, differences in forecast accuracy and reliability between the two training strategies are limited. The monthly batch training set-up exhibits slightly improved performance across most evaluation metrics; however, these gains are modest and do not materially alter the conclusions of the study. The primary motivation for the monthly batch approach is its closer alignment with operational forecasting practice, in which models are periodically re-trained as new data become available, rather than the pursuit of substantial performance improvements.

\begin{figure}[H]
    \centering
    \includegraphics[width=1.1\linewidth]{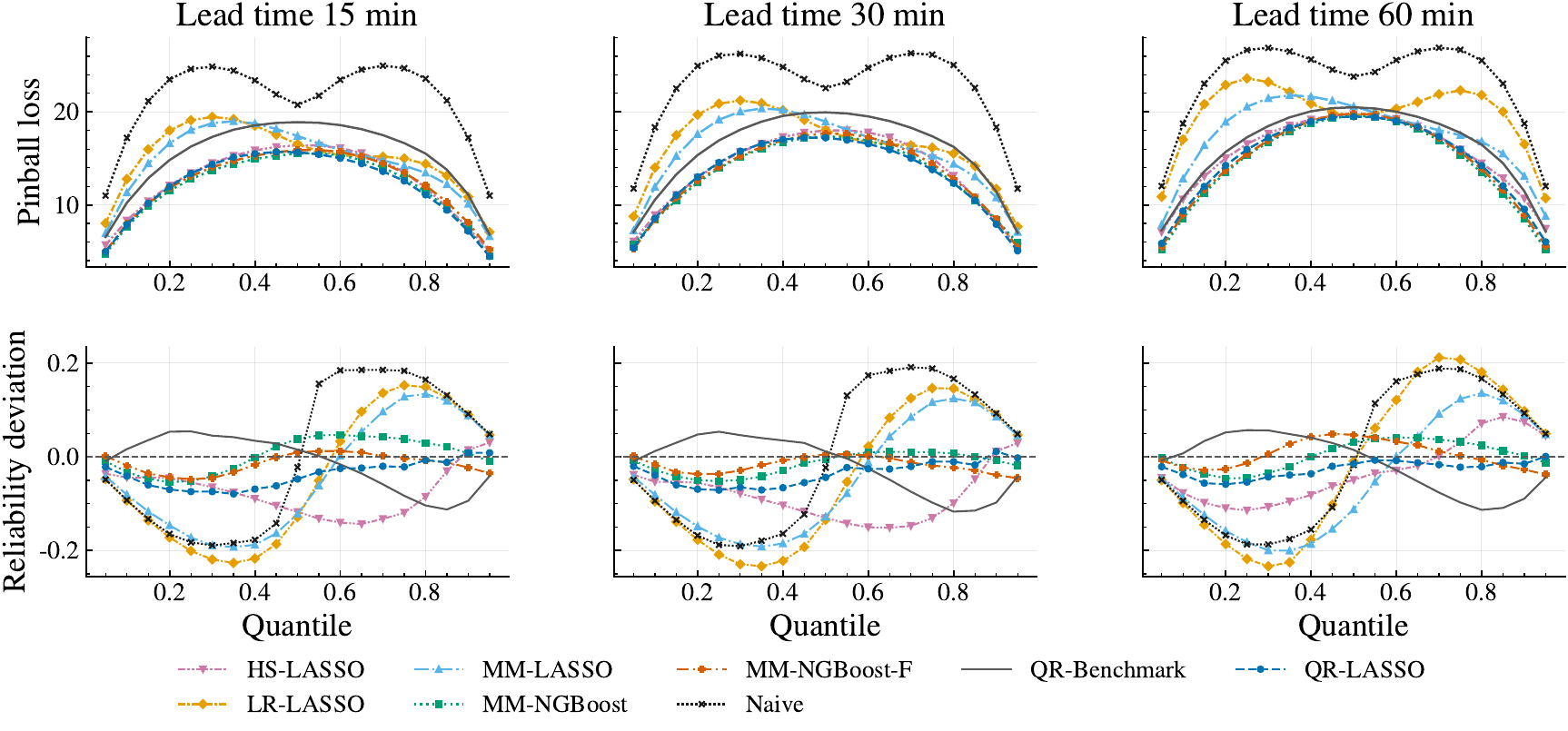}
    \caption{Mean Pinball Loss \eqref{eq: Mean Pinball Loss} and Reliability Deviation \eqref{eq: reliability} at key lead times for each model are evaluated. The results follow a setup where predictions are made for whole 2023, while training set is fixed and  also validation set is fixed for the entire year of 2022. We evaluate our probabilistic forecasts using the entire year of 2023. QR-LASSO and MM-NGBoost, which account for heteroskedasticity, show consistent accuracy against the QR-Benchmark for every lead time, while LR-LASSO and MM-LASSO, which both assume constant variance, perform poorly. \red{MM-NGBoost-F, the fundamentals-only variant of MM-NGBoost, is also shown; see Section~\ref{sec: model performance} for discussion.}
    \label{fig:mpl-rdl_fixed results}}
\end{figure}

\begin{figure}[H]
    \centering
    \includegraphics[width=1.1\linewidth]{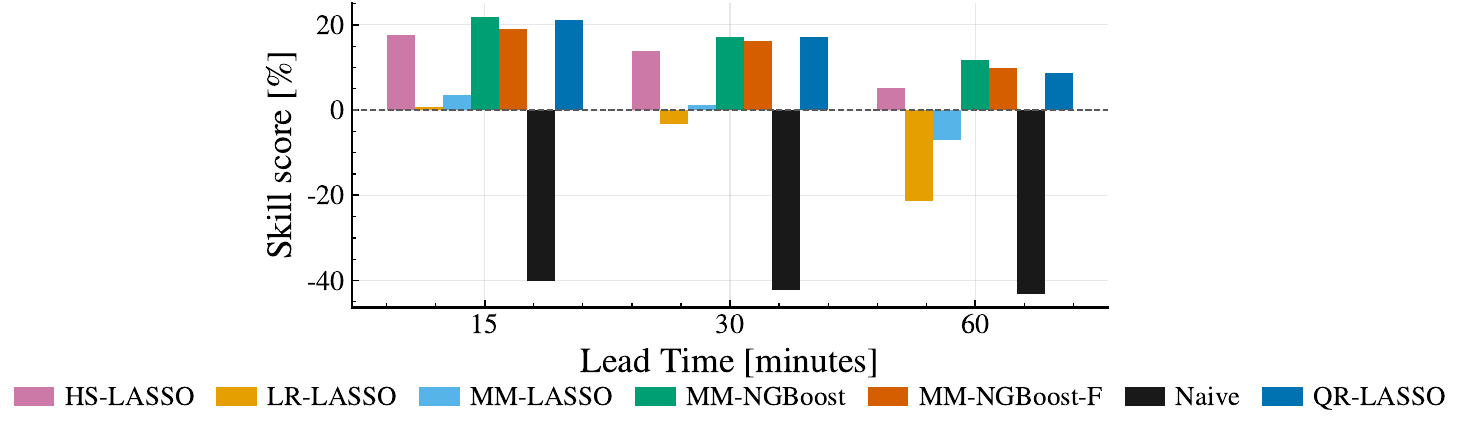}
    \caption{Mean Pinball Loss \eqref{eq: skill score} was used as a skill score to compare our tested models with the benchmark at each key lead time. The levels correspond to the fixed forecasting setup. \red{With the exception of MM-NGBoost-F, which is trained on the same fundamentals-only feature set as QR-Benchmark,} all models presented here utilise information from the intraday market whereas the reference method, QR-Benchmark, does not. 
    The skill score for QR-LASSO represents the gain from incorporating features derived from intraday trades as QR-LASSO and QR-Benchmark differ only in the inclusion of these features.
    \label{fig:skill score_fixed}}
\end{figure}

\end{document}